\documentclass{article}
\usepackage{amsmath,amssymb,color}
\usepackage{feynarts}
\usepackage{epsfig}
\newcommand{\nonesep}{}
\newcommand{\tmbsl}{\ensuremath{\backslash}}
\newcommand{\tmem}[1]{{\em #1\/}}
\newcommand{\tmhlink}[2]{{\color{blue} #1}}
\newcommand{\tmname}[1]{\textsc{#1}}
\newcommand{\tmop}[1]{\ensuremath{\operatorname{#1}}}
\newcommand{\tmstrong}[1]{\textbf{#1}}
\newcommand{\tmtexttt}[1]{{\ttfamily{#1}}}
\newenvironment{itemizedot}{\begin{itemize} }{\end{itemize}}

\begin{document}

\begin{titlepage}

\begin{flushright}
{\bf IFJPAN-IV-2006-4}
\end{flushright}

\vspace{1mm}

\begin{center}
{\LARGE {\tmname{Effective}} \bf 1.0: An Analytic Effective Action Analysis
Library$^*$}
\end{center}
\vspace{5mm}
\begin{center}
{\large \bf James P. J. Hetherington$^a$} {\rm and} {\large \bf Philip Stephens$^{b \dag}$}\\
\vspace{4mm}
{\em $^a$CoMPLEX, Department of Mathematics, University College London,\\
Gower Street, London WC1E 6BT, UK} \\ \vspace{2mm}
{\em $^b$Institute of Nuclear Physics, Polish Academy of Sciences,\\
ul.\ Radzikowskiego 152, 31-342 Cracow, Poland.}
\end{center}

\vspace{15mm}
\begin{abstract}
  {\tmname{Effective}} is a C++ library which provides the user a toolbox to
  study the effective action of an arbitrary field theory. From the field content, gauge
  groups and representations an appropriate action is generated symbolically.
  The effective potential, mass spectrum, field couplings and vacuum expectation 
  values are then obtained automatically; tree level results are obtained analytically 
  while many tools, both numeric and analytic, provide a variety of approaches to deal 
  with the one-loop corrections. This article
  provides a guide for users to who wish to analyze their own models using {\tmname{Effective}}.
  This is done by presenting the code required and describing the physics assumptions behind
  the code. The library can be extended in many ways and discussion of 
  several such extensions is also provided.
\end{abstract}
\vspace{8mm}
\vspace{20mm}
\begin{flushleft}
{\bf IFJPAN-IV-2006-4}
\end{flushleft}

\vspace{5mm}
\footnoterule
{\footnotesize
$^\star$This work is partly supported by the EU grant MTKD-CT-2004-510126
in partnership with the CERN Physics Department and by the Polish Ministry of Scientific
Research and Information Technology grant No 620/E-77/6.PRUE/DIE 188/2005-2008.\\
$^\dag$Correspondence email: pstephens@annapurna.ifj.edu.pl}
\end{titlepage}

{\tableofcontents}

\section{Introduction}

The effective action of a model characterizes many of its important features.
Using the effective action one is able to study the loop corrections and
renormalization group of a model, the nature of spontaneous symmetry breaking
and the mass spectrum~\cite{WeinbergII}. This makes the effective action a powerful object
to investigating phenomenological aspects of a model.

It is generally agreed that there will be new physics beyond the Standard
Model (BSM). The exact nature of this physics is still unknown and many ideas
exist about its nature (see for example~\cite{MartinPrimer} or~\cite{Appelquist:2000nn}). 
In order to validate these
new ideas, one must show that the tree level and one-loop results of this
model agree with our current knowledge of the Standard Model. Additionally,
new particles, their mass spectrum, couplings and symmetry breaking mechanisms must be
elucidated. Often one wishes to only slightly deviate from an existing model by
introducing a new field or a new interaction. Such deviations can often create
large differences in the structure of the model, particularly at one-loop. 
Existing codes for determining
the spectra of models such as SOFTSUSY~\cite{Allanach:2001kg}, ISASUGRA~\cite{Baer:1999sp}
and SUSPECT~\cite{Djouadi:2002ze} do so for a particular model and are not easily extensible
to study models with a changed field structure. We believe that this flexibility
will be important in the post-standard-model era, as it will be more important
to quickly get an approximate analysis of a new model, rather than to develop
a very accurate code for a well understood model.

Presented here is a new C++ library which provides a toolbox for the study of
an arbitrary model. The code has been designed with the flexibility to allow a user
to change the tools in the toolbox to suit their own purposes.

\subsection{What is {\textsc{Effective}}?}

In order to facilitate exploration of new physics an analytic tool has been created
that automates many of the various uses of the effective action. The tool operates
analytically when possible but reverts to numerics when necessary. Our code typically
treats the tree-level action analytically while evaluating loop corrections numerically.
This way the analytic structure of the tree level action can be derived and the
consequences of the addition of new terms studied. The effect of symmetry breaking mechanisms on
the physical masses and couplings of the action can be studied numerically in
the same framework.

Our solution is presented as a C++ library {\tmname{Effective}}. This
library provides a toolbox for the user to study an action and variants of
that action. {\tmname{Effective}} can be used to study many models with interesting
phenomenological features. Examples include reggeized gluons~\cite{Antonov:2004hh}
and R-parity violating SUSY~\cite{Barbier:1998fe,JamesThesis}. It is possible to 
build extensions to the library to study virtually any model which can be written on 
paper. This code does not provide a simple executable to study an arbitrary model. 
Instead, a user must build their model with the {\tmname{Effective}} library and create 
an executable that probes the desired physics of the model.
The code can be freely obtained from the {\tmname{Effective}} web-page which
is currently at the URL:
{\tmhlink{http://stephens.home.cern.ch/stephens/effective}{}}. Installation
instructions and full code documentation may also be found there.

\subsection{Library Features}

{\tmname{Effective}} has been written in object-oriented C++. This code has
been designed for extensibility from the outset. The possible
extensions will be discussed in section \ref{sec:custom}. The code is built
on the {\tmname{GiNaC}} algebraic engine~\cite{GiNaC}; as {\tmname{GiNaC}} 
is provided as a C++ library our code
is seamlessly integrated with the analytic engine. Not only does this
reduce computational overhead but allows for the implementation of many complicated
routines, many of which would be extremely cumbersome in the languages built
into proprietary analytic software.

This code was initially developed to provide some of the same features found in SUSY
spectrum generators such as {\tmname{SOFTSUSY}}~\cite{Allanach:2001kg}, 
{\tmname{ISASUGRA}}~\cite{Djouadi:2002ze} and {\tmname{SUSPECT}}~\cite{Baer:1999sp}. 
As such this tool can automatically generate the
one-loop mass spectrum of a model. The parameters can be run between different
scales according to the renormalization group equations (RGE) in order to give
a consistent parameter set. In this version, the RGEs need to be provided; a
future extension to generate the one-loop RGEs automatically is planned.

It is worth noting that this code is not intended to replace the much more
efficient and accurate codes like {\tmname{SOFTSUSY}}, {\tmname{ISASUGRA}} and
{\tmname{SUSPECT}}. Where dedicated code for a model exists, it will be faster and
more reliable. Instead, this code can mimic the physics present in those
codes, as well as many other actions and symmetry breaking mechanisms. 
{\tmname{Effective}} should be used to pioneer the study of a new model, with 
dedicated code for the model being written when it is clear it is interesting
enough for a more precise analysis to be worthwhile.

\subsection{Organization}

This article is structured in the following way. In the first section the field
content of a model is discussed. The default properties and interactions of
the fields implemented in {\tmname{Effective}} are given and the routines
needed to include the desired fields in a users model are presented. Also, the
discussion about the vacuum expectation value (VEV) of a field is given in
this section.

The user can also implement additional interaction terms, such as Yukawa
couplings. The routines used to specify these are discussed in section
\ref{sec:interaction}. For SUSY models, specification of the superpotential 
will imply further couplings between the fields. Once these three features 
are implemented the effective action can be derived.

Next, we discuss how {\tmname{Effective}} can be used to analyze the model 
specified. The first step in this process is the generation of the effective 
potential from the action. This is minimized to obtain the vacuum expectation 
values (VEV). This process is given in section~\ref{sec:effpot}. Next comes 
determination of the mass matrices and field couplings. The appropriate calls 
to the library are discussed in section~\ref{sec:massm} along with the 
routines used to access the mass matrices and mixing angles. The one-loop 
corrections to these matrices and mixing angles are also discussed in
this section. The last feature of building a model is the RGEs. These are
discussed in section \ref{sec:rge}. We then present the way in which the 
specification of additional observables which one wishes to calculate using 
{\tmname{Effective}} can be carried out by defining appropriate Feynman 
diagrams.


All of the required routines in sections~\ref{sec:field}-\ref{sec:rge} will 
be discussed in terms of the electro-weak model. The full definition of this 
model will be given in the appendix and two examples using this model to 
produce some physical results will be given in section \ref{sec:ew}.

Section \ref{sec:custom} is reserved to discuss some of the various ways in which a
user can define their own classes to replace the default ones in
{\tmname{Effective}}. These customizations allow the user the possibility of
implementing almost any feature they would like. Unfortunately, to customize
the code, one must have strong C++ knowledge as well as a good understanding
of the {\tmname{GiNaC}} engine. This section will be quite technical and as
with most programming, to truly appreciate the content one must try to
implement things for themselves. To that end more programming oriented
tutorials can be found on the website.

\section{Field Content\label{sec:field}}

Definition of the field content of a model consists of several parts. First
one must define the gauge groups that define the interactions of the fields.
The group structure is defined by the class \tmtexttt{GaugeGroup}. 
Once the group has been specified we define the model's field content. There 
are two types of fields. The gauge bosons or gauge supermultiplet is a 
mediator of the interactions described by the gauge group; 
these are provided by the class \tmtexttt{GaugeField}. The matter fields 
interact with the \tmtexttt{GaugeField}s and with each other based on 
representations of the gauge group. These are described by the class 
\tmtexttt{MatterField}. Once the \tmtexttt{GaugeField}s and 
\tmtexttt{MatterField}s are defined, one specifies for 
which scalar fields the effective potential will be analyzed to determine 
whether they will have non-zero VEVs.

It must be noted that the class \tmtexttt{Field} and its subclasses represent
a set of fields which share the same properties. For example the colour octet of
gluon fields are all contained in one \tmtexttt{GaugeField} object.

\subsection{Gauge Groups}

{\tmname{Effective}} has three subclasses of \tmtexttt{GaugeGroup} defined. These
are \tmtexttt{U1Group}, \tmtexttt{SU2Group} and \tmtexttt{SU3Group}. Other
groups could be implemented. This is discussed in section \ref{sec:custom}.

The \tmtexttt{GaugeGroup}s of a model are defined in the user supplied
routine \tmtexttt{void createGaugeGroups()}. In the electro-weak model 
we have the group structure $SU(2) \times U(1)$, to specify this we
write
\begin{verbatim}
void ElectroWeak::createGaugeGroups() {
  addGaugeGroup(new U1Group("U1", "{g'}", this, U1b));
  addGaugeGroup(new SU2Group("SU2", "{g_W}", this, SU2w));
}
\end{verbatim}
This code block shows how to create the \tmtexttt{GaugeGroup}s and include
them into the model. The function \tmtexttt{addGaugeGroup()} adds a pointer to
a \tmtexttt{GaugeGroup} into the model. The constructor for the groups takes 4
arguments. First is a name which the group will be referenced by; this name
also doubles as the plain text label for the coupling of the group. The second
argument is the LaTeX name of the group's coupling parameter. The third
argument is a pointer to the model which this group belongs to and the last
argument is an integer number unique to the group. This is referred to as the
line of the group. This allows multiple groups with the same group structure,
while still keeping the properties of the groups independent of each other.

\subsection{Gauge Fields}

Now that the gauge groups are defined, the gauge fields can be given. These
are fields which mediate the interaction defined by the group. The kinetic
terms of the effective action for the fields are defined by the spin of the field
and whether it has a supersymmetric partner. In {\tmname{Effective}} there are
three default spin classes implemented. These are \tmtexttt{VectorSpin},
\tmtexttt{FermionSpin} and \tmtexttt{ScalarSpin}. In {\tmname{Effective}} a
\tmtexttt{GaugeField} with \tmtexttt{VectorSpin} has the following kinetic
term
\begin{eqnarray}
  \mathcal{L}_{\tmop{kin}}^{\tmop{GF} (V)}  & = & - \frac{1}{4} F^{a \mu \nu}
  F_{\mu \nu}^a,  \label{eqn:GFVkin}
\end{eqnarray}
where the sum over $a$ is implied and
\begin{eqnarray}
  F^a_{\mu \nu} & = & \partial_{\mu} A^a_{\nu} - \partial_{\nu} A^a_{\mu} + g
  f^{a b c} A^b_{\mu} A^c_{\nu} . 
\end{eqnarray}
In these equations the indices are in the adjoint representation and $f^{a b c}$
is the structure constant of the group, $A_{\mu}^a$ is the vector field and
$g$ is the coupling constant of the group. As we wish to support $N=1$ SUSY, we
also provide for spinor fields in the adjoint representation. If the 
\tmtexttt{GaugeField} is a fermion field, the kinetic term is
\begin{eqnarray}
  \mathcal{L}^{\tmop{GF} (F)}_{\tmop{kin}} & = & -i \bar{\sigma}^{\mu} 
  \lambda^{\dag a} D_{\mu}^{a b} \lambda^b, 
\end{eqnarray}
with
\begin{eqnarray}
  D_{\mu}^{a b} & = & \delta^{a b} \partial_{\mu} + i g f^{c a b} A_{\mu}^c . 
  \label{eqn:GFFcov}
\end{eqnarray}
Again, $g$ is the coupling constant of the group, $f^{a b c}$ is the structure
constant of the group and $A_{\mu}^a$ is the vector field which mediates the
interaction. $\lambda^a$ is the gauge fermion and $\bar{\sigma}^{\mu}$ are the spin matrices
\begin{eqnarray}
  \bar{\sigma}^0 =& \left( \begin{array}{cc}1&0\\0&1\end{array} \right),\,\,\,\,
  \bar{\sigma}^1 &= \left( \begin{array}{cc}0&-1\\-1&0\end{array} \right),\nonumber \\
  \bar{\sigma}^2 =& \left( \begin{array}{cc}0&i\\-i&0\end{array} \right),\,\,\,\,
  \bar{\sigma}^3 &= \left( \begin{array}{cc}-1&0\\0&1\end{array} \right). \nonumber
\end{eqnarray}

In {\tmname{Effective}}, fermions are treated as Weyl fermions
and the $\bar{\sigma}^{\mu}$ object is merely a placeholder representing the spin
structure. If one wanted to code the action in terms of Dirac fermions, this could 
also be done. 

{\tmname{Effective}} by default has only implemented the terms which
correspond to the N=1 SUSY case. As such, there is no implementation of a
gauge field with scalar spin. This is an extension that could be easily added,
however.

Now to see how the $B$ and $W$ bosons are added to the electro-weak model we
provide the implementation of the \tmtexttt{void createGaugeFields()} routine.
\begin{verbatim}
void ElectroWeak::createGaugeFields() {
  VectorSpin v;
  addField("B",new GaugeField("B", "B", v, getGaugeGroup("U1")));
  addField("W",new GaugeField("W", "W", v, getGaugeGroup("SU2")));
}
\end{verbatim}
From this code we can see that a field is added to the model with the
\tmtexttt{addField()} routine. The first argument is a string reference to the
field and the second argument is a pointer to a \tmtexttt{Field}. The
\tmtexttt{GaugeField} constructor takes four arguments. First is the plain
text label of the field, followed by the LaTeX label. The third argument is
the spin type of the field, in this case both fields are \tmtexttt{VectorSpin}
types. Lastly is a pointer to the \tmtexttt{GaugeGroup} that this field is a
mediator of. The routine \tmtexttt{getGaugeGroup()} returns the
\tmtexttt{GaugeGroup} pointer referenced by the string.

If one wanted to set a superpartner for the field this is done in the
constructor. For example if we now wanted to add the fermionic superpartner to
the $B$ boson, this would be done by
\begin{verbatim}
FermionSpin f;
addField("Bino", new GaugeField("Bino", "\\tilde{B}", f,
                                getGaugeGroup("U1"),
                                getGaugeField("B"));
\end{verbatim}
The arguments are the same as before except we have added a reference to the
superpartner of $\tilde{B}$, $B$. We don't need to give the $\tilde{B}$ as an
argument when we create $B$ because the relationship is set for both of them 
when it is given to $\tilde{B}$. This is also necessary as you can't give a pointer 
to an object which you haven't created yet!

Once the fields are created and added to the model, the terms given above are
automatically entered into the effective action.

\subsection{Matter Fields}

We have now created the gauge fields. We need to create the remaining fields
in the theory which interact with the gauge fields and each other. The class
which defines these fields is \tmtexttt{MatterField}. Again, these fields take
a \tmtexttt{Spin} class in order to define them. The implementation in
{\tmname{Effective}} does not yet provide for a \tmtexttt{VectorSpin} type of
\tmtexttt{MatterField}. The kinetic terms for the \tmtexttt{FermionSpin} spin
type of \tmtexttt{MatterField} are
\begin{eqnarray}
  \mathcal{L}^{\tmop{MF} (F)}_{\tmop{kin}} & = & -i \bar{\sigma}^{\mu} 
  \bar{\psi}^{\{A\}} D^{\{A\}, \{B\}}_{\mu} \psi^{\{B\}},  \label{eqn:mffkin}
\end{eqnarray}
where the sums over the sets $\{A\}$ and $\{B\}$ are implicit and
\begin{eqnarray}
  D_{\mu}^{\{A\}, \{B\}} & = & \delta^{\{A\}, \{B\}} \partial_{\mu} + i
  \sum_{i \in \{G\}} e_i g_i t^{a_i}_{A_i B_i} A^{a_i}_{\mu} \delta^{\{A'_i
  \}, \{B'_i \}} .  \label{eqn:covferm}
\end{eqnarray}
In these equations $\{A\}$ and $\{B\}$ represent the set of indices from the
fundamental representation for all of the groups the \tmtexttt{MatterField}
interacts with. The sets $\{A'_i \}$ and $\{B'_i \}$ are the sets of remaining
indices when index $i$ is removed. The sum in eqn. (\ref{eqn:covferm}) is over
all of the gauge groups, $\{G\}$, and $t^{a_i}_{A_i B_i}$ is the generator of
group $i$, $g_i$ is its coupling and $e_i$ is the charge of the field in
the group. $A^{a_i}_{\mu}$ is the vector mediating the interaction for group
$i$. Similarly the kinetic term for the \tmtexttt{ScalarSpin} can be written
as
\begin{eqnarray}
  \mathcal{L}^{\tmop{MF} (S)}_{\tmop{kin}} & = & \left( D^{\{A\}, \{B\}}_{\mu}
  \varphi^{\{B\}} \right)^{\dag} \left( D^{\mu \{A\}, \{B\}} \varphi^{\{B\}}
  \right),  \label{eqn:mfskin}
\end{eqnarray}
where the covariant derivative is the same as in the \tmtexttt{FermionSpin}
case and the sum over the sets $\{A\}$ and $\{B\}$ are implicit.

If a \tmtexttt{MatterField} has a superpartner, there is an additional
interaction which is
\begin{eqnarray}
  \mathcal{L}^{\tmop{MF} (F)}_{\tmop{SUSY}} & = & \sum_{i \in \{G\}} \frac{e_i
  g_i}{\sqrt{2}} t^{a_i}_{A_i B_i} \left[ (\varphi^{\dag})^{A_i, \{A'_i \}}
  \psi^{B_i, \{A_i' \}} \lambda^{a_i} + \bar{\psi}^{A_i, \{A_i' \}}
  \varphi^{B_i, \{A_i' \}} \lambda^{\dag a_i} \right] \nonumber\\
  &  & + c.c. 
\end{eqnarray}
Here $\lambda^{a_i}$ is the gauge fermion of group $i$, $\varphi$ and $\psi$
are supersymmetric partners, where $\varphi$ is a scalar and $\psi$ is a
fermion. There is one additional term in a SUSY theory. This is
\begin{eqnarray}
  \mathcal{L}^{\tmop{MF} (S)}_{\tmop{SUSY}} & = & \sum_{i, j} \sum_{k \in G_{i
  j}} e_i e_j g_k \left( \left( \varphi_i^{\dag} \right)^{A_k, \{A_k' \}}
  t^{a_k}_{A_k B_k} \varphi_i^{B_k, \{A'_k \}} \right) \nonumber\\
  &  & \times \left( \left( \varphi_j^{\dag} \right)^{A_k, \{A_k' \}}
  t^{a_k}_{A_k B_k} \varphi_j^{B_k, \{A'_k \}} \right), 
\end{eqnarray}
where again the sum over $a_i$, $A_k, B_k$ and $\{A_k' \}$ is implicit. $G_{i
j}$ is the set of groups field $i$ and field $j$ have in common. In order to
include these terms, all scalars of the theory must be defined. These are
added when the model initializes. If the user calls the routine
\tmtexttt{Model\tmtexttt{::noDterms()}}, then these terms will not be added,
e.g. \tmtexttt{ElectroWeak::noDterms()}. This must be called in the
constructor of the model.

We now give an example of adding leptons and a $S U (2)$ Higgs field to the
electro-weak model. This is provided by the implementation of the
\tmtexttt{void createMatterFields()} routine.
\begin{verbatim}
void ElectroWeak::createMatterFields() {
  numeric half(1,2);
  ScalarSpin s;
  FermionSpin f;
  addField("l", new MatterField("l", "\\ell", f, famsize,
                              getGaugeGroup("U1"), -half,
                              getGaugeGroup("SU2"), 1));
  addField("eR", new MatterField("eR", "e_R", f, famsize
                               getGaugeGroup("U1"),-1));
  addField("H", new MatterField("H","H",s,1,getGaugeGroup("U1"),
                              half, getGaugeGroup("SU2"),1));
  // Implement VEV code here
}   
\end{verbatim}
We see that the creation of a \tmtexttt{MatterField} is similar to that of
the \tmtexttt{GaugeField}. The difference is that the \tmtexttt{MatterField}
can have several groups and a different charge under each group. If we look at
the arguments for the \tmtexttt{MatterField} constructor we see that the first
three arguments are the same as for the \tmtexttt{GaugeField}. These are
followed by an integer representing the number of families of the field,
followed by pairs of \tmtexttt{GaugeGroup} pointers and charges. The 
\tmtexttt{MatterField} class allows each matter field to be a representation
of no more than 4 gauge groups. If
more than 4 groups are needed for a field, then a new implementation of
\tmtexttt{MatterField} must be made.

Similar to the \tmtexttt{GaugeField} the superpartner can be set by adding an
additional argument to a \tmtexttt{MatterField} pointer. This can be retrieved
by a call to \tmtexttt{getMatterField()} with the appropriate string as an
argument. As in the \tmtexttt{GaugeField} this superpartner needs to be passed
only to the constructor of the second field of the pair. The appropriate
relation is set for both fields.

We can also see from the previous code segment that the fields $\ell$ and
$e_R$ are defined as the electron fields (and all families), not the positron
field. This can be seen as the charge of the $e_R$ field is $- 1$, and the
charge of the $\ell$ field is $- \frac{1}{2}$ under $U (1)$ group and $1$
under the $S U (2)$ group. This gives the electron field a $U
(1)_{\tmop{QED}}$ charge of $- 1$.

\subsection{Vacuum Expectation Values}

One of the most important uses of the effective potential is its minimization 
to determine which, if any, of the fields in the model may develop vacuum 
expectation values. However, as it would be computationally prohibitive to do 
this simultaneously for all scalar fields, we use the \tmtexttt{Parameter} object 
to specify which fields will be analyzed for VEVs. 
Once all the field content of the model is defined, the user can give some of
the \tmtexttt{MatterField}s a non-zero vacuum expectation value. This value
can be given as a parameter of the model, whose value can later be changed to
study the impact of the VEV on physical results.

In the previous section we gave the first part of the implementation of the
\tmtexttt{void ElectroWeak::createMatterFields()} routine. We now insert the
VEV in order to complete this routine. The first step is to define a parameter
which the user can use to modify the value of the VEV. This is achieved with
the code
\begin{verbatim}
Parameter upsilon = addParameter("HiggsVev","\\upsilon",220.0,
                                 Parameter::vev);
\end{verbatim}
This code creates a \tmtexttt{Parameter} whose text name is
\tmtexttt{HiggsVev} and the LaTeX name is $\upsilon$. The default value of
this parameter is 220.0. The last argument indicates that this parameter is a
VEV. The function \tmtexttt{addParameter()} adds this parameter to the model.
The numeric value of all parameters is centrally stored in the model. This way
a change to the value universally changes in all references to the parameter.
The parameter can later be retrieved by calling \tmtexttt{Parameter
Model::getParam()} with the string given as an argument. To retrieve the
parameter defined above, the string \tmtexttt{HiggsVev} must be passed.

There is also a routine \tmtexttt{numeric\& Model::getParameter()} which
takes a string as its argument. This routine returns the value (by reference)
of the parameter. The value of this \tmtexttt{numeric} object may be changed
and the change will propagate to all of the parameters, but the
\tmtexttt{numeric} object should not be used when creating expressions. Doing
so will put only the current value of the parameter into the expression.
Future changes to the value will not change the expression created. 
Instead, if the user wants to create an expression with the parameter in it,
they must create the expression with the \tmtexttt{Parameter} object retrieved
by the call to \tmtexttt{Model::getParam()}.

Now the parameter which defines the VEV has been created, the field with the
non-zero VEV can be defined. This is done by the call to the
\tmtexttt{addVev()} routine. In the electro-weak model we want to set the real
part of the $H_2$ field to have a non-zero VEV. This is achieved by
\begin{verbatim}
addVev("HiggsVev",getField("H"), lst(getIndex("H","SU2")==2),
       (upsilon+Model::star));
\end{verbatim}
The first argument is a string which identifies this VEV. Note that this does
not need to be the same as the string which identifies the
\tmtexttt{Parameter} of the VEV, though it also need not be different. The
second argument is a pointer to the field which the VEV is being set for. The
third argument is a \tmtexttt{lst} object from {\tmname{GiNaC}}. This list
specifies what substitutions to make on the \tmtexttt{Field} multiplet to get
the desired field. In this example the $S U (2)$ index, retrieved by calling
\tmtexttt{Model::getIndex()}, is set to $2$. The last argument is the
expression to replace the original field by. The object \tmtexttt{Model::star}
represents the field in question. In this example the substitution
\begin{eqnarray}
  \mathcal{R}(H_2) & \rightarrow & \upsilon +\mathcal{R}(H_2), 
\end{eqnarray}
is made. A \tmtexttt{Parameter} in {\tmname{Effective}} is always real. In
order to replace the imaginary part of a field with a VEV, an additional
argument, \tmtexttt{Imag}, must be given to the \tmtexttt{addVev()} routine.
This would then make the substitution
\begin{eqnarray}
  \mathcal{I}(H_2) & \rightarrow & \upsilon_I +\mathcal{I}(H_2), 
\end{eqnarray}
where $\upsilon_I$ is a new parameter specifying the imaginary part of the
VEV.

The last step is to tell the model that this parameter is a VEV, and not some
other kind of parameter. This is important for the code as the VEV parameters 
are treated internally differently than other parameters. This is achieved by the call
\begin{verbatim}
addVevParameter(upsilon);
\end{verbatim}

\subsection{Default Behaviour of Field and Spin Classes}

{\tmname{Effective}} has many default behaviours built into the classes
described above. These behaviours will be sufficient for most models, however,
there are many models for which the user will need to implement their own
classes. The \tmtexttt{GaugeGroup}, \tmtexttt{Spin\tmtexttt{}} and
\tmtexttt{Field} classes have been designed with enough flexibility to
accomodate most modifications. Details of how one would make such
modifications are reserved for section \ref{sec:custom}. Here we present the
default behaviour of the \tmtexttt{GaugeField},
\tmtexttt{MatterField\tmtexttt{}}, \tmtexttt{ScalarSpin},
\tmtexttt{FermionSpin} and \tmtexttt{VectorSpin} classes.

The \tmtexttt{GaugeField} class is designed to take one \tmtexttt{GaugeGroup}
and a \tmtexttt{Spin} object. This will define the field multiplet of the
given spin. The equations that dictate the kinetic terms where given in eqns.
(\ref{eqn:GFVkin}-\ref{eqn:GFFcov}). This allows one to try all sorts of things,
not all of which will be renormalizable! For example, one could create a scalar
gauge field not part of a supermultiplet, resulting in a kinetic term of the form
\begin{eqnarray}
  \mathcal{L}^{\tmop{GF} (S)}_{\tmop{kin}} & = & \left( D_{\mu}^{a b}
  \varphi^b \right)^{\dag} \left( D^{\mu a b} \varphi^b \right) . 
\end{eqnarray}
In this equation the covariant derivative is given by eqn. (\ref{eqn:GFFcov})
and the indices are in the adjoint representation. Note that this is really a
side-effect of the code. {\tmname{Effective}} has not been designed with such
terms in mind.

The \tmtexttt{MatterField} class has been designed to take a family size, up
to four \tmtexttt{GaugeGroup}s and charges under each group. This then
generates the kinetic terms given in eqns.
(\ref{eqn:mffkin}-\ref{eqn:mfskin}). If a \tmtexttt{VectorSpin} where to be
passed to this class this would cause an error. The \tmtexttt{VectorSpin}
class has currently only been defined to work with \tmtexttt{GaugeField}
classes and subclasses. By implementing a new \tmtexttt{Spin} subclass one
could have a vector field with an arbitrary gauge representation.

The \tmtexttt{Spin} class defines several properties. The first one of
importance is to specify how the field will be handled with respect to CP;
whether the fields of a spin are complex or purely real. If they
are complex the expressions are created so there is a unique expression for
the real and for the imaginary part of a field. This implies
\begin{eqnarray}
  F & = & \frac{\mathcal{R}(F) + i\mathcal{I}(F)}{\sqrt{2}} . 
\end{eqnarray}
As all fundamental terms in the expressions of {\tmname{Effective}} are real,
this substitution allows the software to treat the field $F$ as a complex
field in terms of its real and imaginary parts. When one wants to probe the
action for information on a field, they must therefore ask specifically about
the real and the imaginary part of a field. The default behaviour of the
\tmtexttt{ScalarSpin} and \tmtexttt{FermionSpin} classes is to treat the
fields as complex, while the \tmtexttt{VectorSpin} class treats the field as
real. The division of the real and imaginary parts was intentional in order to
study CP violating terms directly. A future extension will be to allow the
user to decide whether to treat the basic objects in {\tmname{Effective}} as
complex or not.

The last relevant property of the \tmtexttt{Spin} class is whether the spin
type contains a Lorentz index or not. In {\tmname{Effective}} the Lorentz
structure is treated explicitly, while the Dirac structure of the fermions is
implicit. This means that the expressions of vectors have explicit Lorentz
indices attached, while the fermions do not have spinor indices. A future
extension will be to include the spinor indices on the fermions.

\section{Interaction Terms\label{sec:interaction}}

The next step of the creation of the effective action of a model is to include
non-kinetic terms. This includes terms such as the Higgs self interaction and
the Yukawa couplings. In supersymmetric theories the
superpotential replaces the need to implement some of these terms.

\subsection{How to add Desired Expressions}

To include the interaction terms in {\tmname{Effective}} the routine 
\tmtexttt{void addOtherTerms()} must be implemented. As we have seen with the
routines in the previous section, this routine is a virtual routine of the
\tmtexttt{Model} class. Thus the user's model is a subclass of
\tmtexttt{Model} and \tmtexttt{addOtherTerms()} is a routine of their class.

We now discuss the example of adding the Yukawa and Higgs self interaction
terms to the electroweak model, defined by the class \tmtexttt{ElectroWeak}.
We will begin by showing the terms
\begin{eqnarray}
  \mathcal{L}_{\tmop{Higgs}} & = & \mu^2 \varphi_i^{\dag} \varphi_i - \lambda
  \left( \varphi_i^{\dag} \varphi_i \right)^2,  \label{eqn:higgs}
\end{eqnarray}
where there is an implicit sum over $i$. In the second term this sum is
performed before the square operation. This first term is given by
\begin{verbatim}
Parameter mu = addParameter("mu","\\mu", 1.0);
vector<idx*> indices = getField("H")->getIndices();
ex H = getField("H")->expression();
ex a = pow(mu,2)*H.conjugate()*H;
add(Utils::sumIndices(a,indices).expand());
\end{verbatim}

The first line of code creates new \tmtexttt{Parameter} object for the $\mu$
coupling.  This is followed by a new object which requires explanation. 

\subsubsection{Summation}
In {\tmname{Effective}} all of the implicit sums, like the ones in eqn.
(\ref{eqn:higgs}), must be made explicit. In order to achieve this the user
must tell the code which indices to be summed. Implicit in the creation of the
indices is the values a particular index can take (except Lorentz indices, for
which dimensionality is variable and need not be integer).

The second line of code creates a list of all the \tmtexttt{idx} (pointers)
which the field ``H'' contains (not including any potential Lorentz indices).
This list is then passed to the summation routine \tmtexttt{ex
Utils::sumIndices()} in the last line. This summation routine sums the first
argument, \tmtexttt{a}, over all of the indices in the second argument,
\tmtexttt{indices}. In this example the field $H$ has only
one index in the $SU(2)$ fundamental representation. Explicitly the sum routine
performs
\begin{equation}
\tmop{sumIndices()} = \sum_i \mu^2 H_i^\dag H_i,
\end{equation}
where $i$ is an $SU(2)$ index.
Notice that this summation is passed into the function
\tmtexttt{void add()}. This tells the model to add the expression to the
Lagrangian.

\subsubsection{Expansion}
One other note. The {\tmname{GiNaC}} engine may often keep terms grouped in
multiplication, i.e. $(x + y) (z + w)$, where in the Lagrangian, we want terms
as individual products, i.e. $x z + x w + y z + y w$. The \tmtexttt{expand()}
routine ensures we have this expanded result. It is not necessary to call this
command, the {\tmname{Effective}} library will do the transformation
automatically at a later stage. Performing the expand on the small expression
above is much more efficient than waiting for the library to do it later. This
is because the library will only do it once all the terms have been included
and the VEVs set. Thus, it is recommended that this is done before it is added
to the model.
\vspace{5mm}

Now consider the second part of eqn. (\ref{eqn:higgs}). Here we must perform
the summation before squaring the result. This is easily obtained from
\begin{verbatim}
Parameter lambda = addParameter("lambda", "\lambda", 1.0);
ex la = Utils::sumIndices(H.conjugate()*H,indices).expand();
add(-lambda*pow(la,2));
\end{verbatim}
From the previous discussion all of the relevant pieces are already known. We
create a new \tmtexttt{Parameter} and compute the implicit sum. This sum is
squared, multiplied by the coupling and added to the Lagrangian.

\subsubsection{Families}
The remaining term to include in the Lagrangian for the electroweak model is
the lepton Yukawa coupling. This is given by
\begin{eqnarray}
  \mathcal{L}_{\tmop{Yuk}} & = & - Y^e_{i j} e_{R i} H^a  \bar{\ell}_j^a +
  c.c. 
\end{eqnarray}
We see that there is now a sum over families, $i, j$, as well as a sum over
the $S U (2)$ index $a$. These sums will be performed explicitly in the code.

The code below will include the case for any number of families, with a
special case for only one family. In this code the variable \tmtexttt{famsize}
is a global integer which specifies the number of families.
\begin{verbatim}
ex Ye;
if(famsize != 1) Ye = addFamilyMatrix("Ye", "{Y^e}", famsize);
else Ye = addParameter("Ye", "{Y^e}", 1.0);
idx i = Utils::familyIndex(0,famsize);
idx j = Utils::familyIndex(1,famsize);
ex eR = getField("eR")->expression()
ex l;
if(famsize != 1) l = 
         getField("l")->expression().conjugate().subs(i==j);
else l = getField("l")->expression().conjugate();
indices = getField("l")->getIndices();
ex b = -Ye*l*H*eR;
if(famsize != 1) indices.push_back(&j);
ex res = Utils::sumIndices(b,indices).expand();
add(2*Utils::real(res));
\end{verbatim}
The coupling $Y^e$ can either be a square matrix with dimensionality of the number of
families or just a simple \tmtexttt{Parameter} if there is only one family.
The variables \tmtexttt{i} and \tmtexttt{j} are indices in the ``family''
space. This is the same space that the matrix $Y^e$ is defined in. The routine
\tmtexttt{addFamilyMatrix()} creates a matrix which is
\tmtexttt{famsize}$\times$\tmtexttt{famsize}. Each element of the matrix is
associated with a unique \tmtexttt{Parameter} with name \tmtexttt{Yeij} where
\tmtexttt{i} and\tmtexttt{j} are replaced by the particular value of interest,
i.e. \tmtexttt{Ye12}. The first argument is the string by which the matrix is
stored for future retrieval. This argument also doubles as the value printed
for non-LaTeX output.

\tmtexttt{Utils::familyIndex(0,famsize)} is a utility routine in
{\tmname{Effective}} to retrieve index 0, from a predefined list of indices,
with dimensionality \tmtexttt{famsize}. Such predefined lists exist for all
types of indices (e.g. $S U (2)$, family, Lorentz, etc.) so that indices can
be matched and replaced when creating expressions. The
\tmtexttt{familyIndex(i,s)} routine will return the $i$th index from the list
with family dimension \texttt{s}.
When a field is created, it's family index is always the 0th element of that
list (the same applies to the group indices). When the family matrix is
created, it has two indices, the 0th and the 1st. Thus we retrieve the 0th and
the 1st and store them in the variables \tmtexttt{i} and \tmtexttt{j},
respectively, for later use.

\subsubsection{Expression Substitution}
The next line of interest is when the left handed lepton expression is
retrieved. In the case that \tmtexttt{famsize != 1} we see the extra command
\tmtexttt{subs(i==j)}. This is a very powerful, and useful, function in
{\tmname{GiNaC}}. This instruction takes the expression it is called on (in
this case the full complex expression of $\ell$) and substitutes all
occurrences of \tmtexttt{i} with \tmtexttt{j}. Thus the sum over \tmtexttt{i}
does not affect $\ell_j$ and in the sum over \tmtexttt{j}, $e_{R i}$, is
unaffected. In both summations $Y^e_{i j}$ does change. We can see this
summation in the second to last line. We see in the line before that the
family index \tmtexttt{j} is added to the list of summation indices. We must
remember that when we retrieved the indices from the left-handed lepton field
it only contains one family index, \tmtexttt{i}. Thus to do the double
summation, we must add \tmtexttt{j} to our list. Explicitly, this summation is
\begin{equation}
\tmop{sumIndices()} = \sum_a \sum_i \sum_j Y^e_{ij} e^\dag_{Ri} \ell^a_j H_a,
\end{equation}
where $a$ is the $SU(2)$ index.
The function \tmtexttt{Utils::real()} in the last line returns only the real
component of the expression. The factor of 2 is there since $X + c.c. = 2
\tmop{Re} (X)$.

\section{Superpotential}

The most general set of renormalizable SUSY interactions are given by
\begin{eqnarray}
  \mathcal{L}_{\sup}  & = & - \frac{1}{2} W^{i j} \psi_i \psi_j + W_{}^i
  W^{\star}_i + c.c.,  \label{eqn:sup}
\end{eqnarray}
where
\begin{eqnarray}
  W & = & \frac{1}{2} M^{i j} \varphi_i \varphi_j + \frac{1}{6} y^{i j k}
  \varphi_i \varphi_j \varphi_k, \\
  W^{i j} & = & \frac{\delta^2 W}{\delta \varphi_i \delta \varphi_j}, \\
  W^i & = & \frac{\delta W}{\delta \varphi_i} . 
\end{eqnarray}
In these equations the fields $(\psi_i, \varphi_i)$ form a supermultiplet.
This means that $M^{i j}_{^{}}$ is the mass matrix of the fermion fields and
$y^{i j k}$ is the Yukawa coupling of a scalar, $\varphi_{_k}$ to fermions
$\psi_i$ and $\psi_j$. Technically, the superpotential is defined in terms of
supermultiplets. In {\tmname{Effective}}, this is not the case. For the N=1
supersymmetric case, we can implement the superpotential as a function of the
scalar fields. A future enhancement of the code will change this to a true
definition in terms of supermultiplets.

\subsection{Implementation of MSSM Superpotential}

We now give an example of the MSSM superpotential. Just as the previous
section, we will rely on making explicit the implicit summations that occur,
for example, in eqn. (\ref{eqn:sup}). The superpotential is implemented by
defining the function \tmtexttt{ex superPotential()} in user's model subclass.
Note that this function now returns an expression, unlike the
\tmtexttt{addOtherTerms()} routine, the user must not call the
\tmtexttt{Model::add()} routine on terms in the superpotential. We will
implement the MSSM superpotential
\begin{eqnarray}
  W_{\tmop{MSSM}} & = & \varepsilon_{a b} \left( Y^d_{i j}  \bar{d} H_1^a Q^b
  - Y^u_{i j}  \bar{u} H_2^a Q^b + Y^e_{i j}  \bar{e} H_1^a \ell^b \right) +
  \mu H_1 H_2 .  \label{eqn:MSSMpot}
\end{eqnarray}
In fact we will only show the implementation of the $Y^e_{i j}$ term. The
process of implementing the other terms is straightforward. Also note that, as
mentioned before, the superpotential is defined in terms of supermultiplets as
is eqn. (\ref{eqn:MSSMpot}). The implementation will only be defined in terms
of the scalar fields of the supermultiplets.
\begin{verbatim}
ex Ye = addParameter(Ye,Y_e,1.0);
ex superPot = 0;
vector<idx*> sumIndex;
MatterIndex a = getGaugeGroup("SU2")->matterIndex(0);
MatterIndex b = getGaugeGroup("SU2")->matterIndex(1);
idx i = Utils::familyIndex(0,famsize);
idx j = Utils::familyIndex(1,famsize);
ex epsab = epsilon_tensor(a,b);
ex h1 = getField("H1")->expression();
ex h2 = getField("H2")->expression();
ex l = getField("sleptonL")->expression().subs(i==j);
ex eR = getField("sleptonR")->expression();
sumIndex = getField("sleptonL")->getIndices();
sumIndex.push_back(&b);
sumIndex.push_back(&j);
ex temp = Ye*epsij*h1*.subs(a==b)*eR;
superPot += Utils::sumIndices(temp,sumIndex).expand();
// other terms go here...
return superPot;
\end{verbatim}
As before we retrieve the index in the fundamental representation of $S U (2)$
(referred to as the \tmtexttt{MatterIndex}) from a predefined list. We create
the epsilon tensor in this space and the $Y_{i j}^e$ term of eqn.
(\ref{eqn:MSSMpot}) can then be directly implemented as shown. It has been assumed
for this code that the fields \tmtexttt{H1}, \tmtexttt{H2}, \tmtexttt{sleptonL} and
\tmtexttt{sleptonR} have been defined.

\section{Effective Potential\label{sec:effpot}}

The effective potential is one of the main components of {\tmname{Effective}}.
This can be generated at tree-level or at one-loop in {\tmname{Effective}}. \
We give a brief review of the generation of the effective potential in the appendix.
The result in the appendix can be extended to all fields to read, in the
$\overline{\tmop{DR}}$ renormalization scheme,
\begin{eqnarray}
  V_1 & = & \frac{1}{64 \pi^2} \tmop{STr} \left[ M^4  \left( \ln
  \frac{M^2}{\mu^2} - \frac{3}{2} \right) \right] . 
\end{eqnarray}
Here the mass matrix $M$ is the tree-level mass matrix for a set of particles.
The ``supertrace'', $\tmop{STr} [f (X)]$ is defined as
\begin{eqnarray}
  \tmop{STr} [f (X)] & = & \sum_i (2 s_i + 1) (- 1)^{2 s_i} f (X_i), 
  \label{eqn:v1}
\end{eqnarray}
where $s_i$ is the spin of the $i$th particle. This ``supertrace'' is just
the spin weighted trace of the mass matrices.

To find the value of the VEVs, the effective potential can be minimized as a
function of the VEVs. Note that the conventional approach, to find those VEVs 
which set to zero the tadpole diagrams, is equivalent. The tadpole terms are given by
\begin{eqnarray}
  T_i & \equiv & \left. \frac{\partial \mathcal{L}}{\partial \varphi_i}
  \right|_{\{\varphi_j = < \varphi_j >\}}, 
\end{eqnarray}
for all $\varphi_i$ with a non-zero VEV. Thus to solve 
\begin{equation*}
\frac{\partial V_{\tmop{eff}}}{\partial \left< \varphi_i \right>} = 0,
\end{equation*}
minimizes the potential and is equivalent to solving for the zero of the 
tadpole contributions.

\subsection{Effective Potential in {\tmname{Effective}}}

We now turn to routines to access the effective potential in
{\tmname{Effective}}. The effective potential can be accessed with the
\tmtexttt{ex Model::potential()} routine. This routine takes an
\tmtexttt{Approximation::Approximations} flag to specify whether the potential
should be evaluated at tree-level or at one-loop. The default behaviour (if no
\tmtexttt{Approximation::Approximations} flag is given) is to return the
tree-level value. At tree-level, this function returns an analytic result. At
one-loop, it is returned numerically at the point in field-space given by the
current VEVs.  It is also worth noting that the
potential is actually returned with the value of the potential when all VEVs
are zero is subtracted from it, i.e. the difference between the value at the 
current location in field-space and the value at the origin.

To determine the ``natural'' values of the VEVs the effective potential needs 
to be minimized over a set of parameters. This can be achieved by a call to the
\tmtexttt{Numerics::extremizePotential()} routine. This routine takes a list
of \tmtexttt{Parameter}s to minimize the potential over. It also takes an
\tmtexttt{Approximation\-::Approximations} flag to indicate what level of
approximation of the potential to use during the minimization. This routine
uses the direction set (or Powell's) method in multidimensions~\cite{NumericalRecipes} to
minimize the potential. If this method fails, the original values of the
parameters are restored.

There is also a routine, \tmtexttt{exvector Model::tadpoles()} which returns a
list of the values of the tadpoles for the current values of the
\tmtexttt{Parameter}s. This routine also takes an
\tmtexttt{Approximation::Approximations} flag. If none is given the default
behaviour is to return the tree-level result.

Similar to the potential, there is a routine which can scan a parameter set
and find where all all tadpole diagrams are zero. This routine can also use
tree-level or one-loop tadpoles. The routine is
\tmtexttt{Numerics::solveZeroTadpoles()}. This routine takes a list of
parameters to scan over and an approximation to use. This routine also takes a
%
boolean flag which, when true, will save the values of the parameters passed
to the routine and if the routine fails, it will restore the values. If false
is passed, the values of the parameters after a failed call are unpredictable.
This routine uses the Newton-Raphson method of root finding~\cite{NumericalRecipes} to
numerically determine the values of the \tmtexttt{Parameter}s that make all
tadpoles zero.

Both the \tmtexttt{extremizePotential()} and the
\tmtexttt{solveZeroTadpoles()} routines have built in error handling. This
will print to the \tmtexttt{cerr} stream the cause of any failed calls. These
messages are useful to explain odd behaviour of a program after a failed call.
An error logging system will be implemented in a future update so all messages
can be stored in a file.

We now give the example of calling these routines with the
\tmtexttt{ElectroWeak} model that we partially defined in the previous
sections. The full definition of this class can be found in the appendix.
\begin{verbatim}
vector<numeric*> list;
list.push_back(&(ew.getParameter("HiggsVev")));
ex pot = Numerics::extremizePotential(&ew,list);
Numerics::solveZeroTadpoles(&ew,list,true,
                            Approximation::OneLoop);
\end{verbatim}
In this code the variable \tmtexttt{ew} is an instance of the
\tmtexttt{ElectroWeak} class. We see here that the tree-level potential is
minimized in the third line and the value of that minimized potential is
stored in the \tmtexttt{pot} variable. The fourth line then is an example of
finding the value of the Higgs VEV which gives a zero one-loop tadpole.

\section{Mass Matrices\label{sec:massm}}

We now turn to the mass matrices. These are derived from the Lagrangian by
\begin{eqnarray}
  M_{i j} & = & \pm \left. \frac{\partial \mathcal{L}}{\partial \varphi_i
  \partial \varphi_j} \right|_{\{\varphi_k = < \varphi_k >\}}, 
  \label{eqn:massm}
\end{eqnarray}
where $\varphi_i$ represents any field and $< \varphi_k >$ is the expectation
value of field $\varphi_k$. $M_{i j}$ contains the mass-squared matrix for
scalars and vectors, but only the mass matrix for fermions. The plus or minus
term is due to the fact that the mass term for vector bosons have a positive
sign in the Lagrangian, but for fermions and scalars it has a negative sign.

Once we have evaluated all of the elements $i, j$ of the mass matrix we must
diagonalize it. The elements in the Lagrangian are the interaction eigenstates
of the fields. The masses correspond to the mass eigenstates. Diagonalizing
the mass matrix will give us a rotation matrix to mix the interaction states
and give the mass states. The masses of the mass states are then the values of
the diagonalized matrix.

\subsection{Tree Level Mass Matrices}

At tree level the mass matrix generation and diagonalization is
straightforward. Employing eqn. (\ref{eqn:massm}) we can create large matrices
with all of the tree level terms. We can then divide the large matrix into
block diagonal parts. The diagonalization procedure of the large matrix will
give the same results as the diagonalization of the smaller matrices, but is
much more efficient. Since all terms in {\tmname{Effective}} represent real
fields, we have real mass matrices. We can diagonalize the matrices to get
\begin{eqnarray}
  M_{i j} & = & U_{i k} D_{k k} U^T_{k j} . 
\end{eqnarray}
The $D_{k k}$ matrix is a diagonal matrix and the $U_{i j}$ matrix is a
unitary matrix which contains the rotations. From this one finds
\begin{eqnarray}
  D_{k k} & = & U^T_{k i} M_{i j} U_{j k}, 
\end{eqnarray}
and in the Lagrangian this is
\begin{eqnarray}
  \varphi_i^T M_{i j} \varphi_j & = & \left( U^T_{k i} \varphi_i \right)^T D_{k
  k}  \left( U^t_{k j} \varphi_j \right) = \left( \varphi^M_k \right)^T D_{k k} 
  \left( \varphi_k^M \right) . 
\end{eqnarray}
This allows us to interpret $U^T_{k i} \varphi_i$ as the mass eigenstate
$\varphi_k^M$, which implies $U^T_{k i}$ is the mixing angle between interaction
state $\varphi_i$ and mass state $\varphi_k^M$.

In {\tmname{Effective}} the classes \tmtexttt{Mass} and \tmtexttt{MixingAngle}
provide an interface to this process. These objects are {\tmname{GiNaC}}
objects which means they can be directly used in analytic expressions. When
one evaluates these, the appropriate diagonalization is performed and the
numerical element is returned. \

A small discussion is needed about fermion masses. For the vector and scalar
masses the diagonal matrix contains the mass squared. A negative mass-squared
represents a tachyon and in turn has some implications to the parameters and
structure of the theory. For fermions, however, a negative diagonal element is
a negative mass. The mass-squared is still positive so it is not a tachyon.
Instead, an appropriate shift of phase is required. Such a shift of phase
converts the mass of the physical state from negative to positive. In
{\tmname{Effective}}, such a rotation is performed internally automatically.
Thus, fermionic mass states always have positive mass. This can be seen by
shifting the physical mass state by
\begin{eqnarray}
  \varphi^M  & \rightarrow & e^{i \pi / 2} \varphi'^M, 
\end{eqnarray}
This shift leads to
\begin{eqnarray}
  m \varphi^M \varphi^M & = & m \left( i \varphi'^M \right) \left( i
  \varphi'^M \right) = - m \varphi'^M \varphi'^M . 
\end{eqnarray}
As we can see, the mass can be shifted from negative to positive simply by an
appropriate phase shift. This shift amounts to reinterpreting the mass
eigenstate to include the rotation by $U$ and the phase shift.

\subsection{One-Loop Mass Matrices}

When we move to the one-loop corrections to the mass matrices things get
slightly more complicated. Now our Lagrangian couplings are not the bare
parameters of the theory, but instead renormalized ones.

An element of the renormalized Lagrangian would read
\begin{eqnarray}
  \mathcal{L} & = & Z_i \partial_{\mu} \varphi_i \partial^{\mu} \varphi_i -
  \frac{1}{2} Z^M_{i j} M_{i j} Z_i^{1 / 2} Z^{1 / 2}_j \varphi_i \varphi_j, 
\end{eqnarray}
where $Z_i^{1 / 2} \varphi_i$ is the renormalized field $\varphi_{R i}$, and
$Z^M_{i j} M_{i j}$ is the renormalized mass, $M_{R i j}$. The diagonalization
and mixing of states proceeds in the same manner as before. The difficulty now
lies in automatically computing $M_{R i j}$ from the tree-level Lagrangian and 
imposing the renormalization conditions.

\begin{figure}[h]
  \centering
  \input{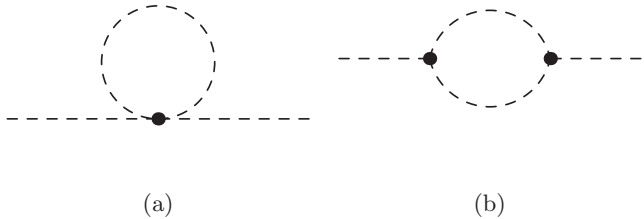}
  \caption{The two topologies for one-loop two point Green
  functions.\label{fig:twopoint}}
\end{figure}

We now define the two point Green functions $\Gamma^{(2)}_{i j} (p^2)$ that
are included in {\tmname{Effective}} at one-loop. At one-loop there are two
topologies that can enter the two point Green functions. These are shown in
figure \ref{fig:twopoint}. We will refer to figure \ref{fig:twopoint}a as a
three-point diagram (as the couplings are three-point couplings), and figure
\ref{fig:twopoint}b as a four-point diagram. The three-point diagrams are
functions of the masses of the two internal lines while the four-point
diagrams are only functions of the mass of the one internal line. All diagrams
are renormalized at a scale $\mu$.

We can now define the values of these individual diagrams for each combination
of spins that can arise. In {\tmname{Effective}}, certain assumptions about
the Lorentz structure of the couplings is made. In models where these
assumptions break down, the one-loop corrections of {\tmname{Effective}} will
be wrong. We have also implemented the contributions in the Feynman gauge.
This means we also must include the appropriate ghost contributions. We will
now discuss the contributions to three sets of diagrams: scalar-scalar,
fermion-fermion and vector-vector. All of the corrections are expressed in
terms of Passarino-Veltman~\cite{Passarino:1979jh} functions in $d$ dimensions. 
A list of the Passarino-Veltman functions and the one-loop mass corrections
are given in the appendix. It is extremely important to note that by including
the complete set of one-loop diagrams for arbitrary couplings and spins we
remain model-independent. However, we restrict ourselves to one-loop diagram
topologies with up to four-point interactions. To include a model which contains
higher-than-four-point interactions one would have to include some more integrals.

\subsection{Renormalization Prescription}

We now turn to the mechanism by which the two-point Green functions are used
to provide the one-loop mass corrections in {\tmname{Effective}}. In
{\tmname{Effective}} the $\overline{\tmop{MS}}$ renormalization prescription
is used. This means that the terms with $(\bar{\epsilon})^{-1}$ from the
Passarino-Veltman (PV) functions are absorbed into the definition of the bare
parameters. In the current version the PV functions are only defined in terms
of their finite contributions. Thus, even though the Green functions are
defined in $d$ dimensions, using $d = 4 - 2 \epsilon$ will not properly give
finite contributions from terms like $\epsilon B_0 (p^2, m_1^2, m_2^2 ;
\mu^2)$. This is equivalent to using the $\overline{\tmop{DR}}$ scheme, where
the momentum are taken in $d$ dimensions, but the vectors and Dirac $\gamma$
matrices are treated in $4$ dimensions. Using the $\overline{\tmop{DR}}$
scheme we define the one-loop renormalized Green functions
\begin{eqnarray}
  \Gamma_R^{(2)} (p^2) & = & \Gamma_{\tmop{tree}}^{(2)} (p^2) +
  \Gamma^{(2)}_{1 - \tmop{loop}} (p^2) . 
\end{eqnarray}
This is equivalent to using counterterms in the Lagrangian. For example the
tree-level Green function for scalar fields is given by
\begin{eqnarray*}
  \Gamma^{(2)}_{\tmop{tree}} (p^2) & \propto & 
  [\partial_{\mu} \varphi]^\dag \partial^{\mu} \varphi - m^2_{0} \varphi^\dag
  \varphi \longrightarrow \left( p^2 - m_{0} \right)
  \varphi^\dag \varphi .
\end{eqnarray*}
Now the one-loop term is of the form
\begin{eqnarray*}
  \Gamma^{(2)}_{\tmop{one} - \tmop{loop}} (p^2) & \propto & \left( A  p^2 + B \right) 
  \varphi^\dag \varphi,
\end{eqnarray*}
so the renormalized term is
\begin{eqnarray}
  \Gamma^{(2)}_R (p^2) & \propto & (1 + A) [\partial_{\mu} \varphi]^\dag
   \partial^{\mu} \varphi - (m_{0} - B) \varphi^\dag \varphi . 
\end{eqnarray}
If one uses the on-shell prescription we require the pole of the propagator to
be equal to the physical mass. This is
\begin{eqnarray}
  \left. \Gamma^{(2)}_R (p^2) \right|_{p^2 = m^2_{R}} & = & 0, 
  \label{eqn:onshell1}\\
  \left. \frac{\partial \Gamma^{(2)}_R (p^2)}{\partial p^2} \right|_{p^2
  = m^2_{R}} & = & 1.  \label{eqn:onshell2}
\end{eqnarray}
This is equivalent to requiring that the renormalized Lagrangian takes the
form
\begin{eqnarray*}
  \mathcal{L}_R & = & \partial_{\mu} \varphi_R \partial^{\mu} \varphi_R +
  m_R^2 \varphi_R \varphi_R .
\end{eqnarray*}

We consider the case where there is mixing between fields; the on-shell
renormalization condition is equivalent to requiring the denominator of the
propagator to vanish. Thus the solution $p^2 = m_R^2$ gives

\begin{equation}
{\rm Det} \left[ \delta_{ij}(p^2 - m^2_{0,ij}) + \Pi(p^2)_{ij} \right] = 0.
\label{eqn:dets}
\end{equation}
For fermion fields this takes the form
\begin{equation}
{\rm Det} \left[ \delta_{ij} (\not{p} - m_{0,ij}) + \Sigma(\not{p})_{ij} \right] = 0.
\label{eqn:detf}
\end{equation}
where $\Pi(p^2)$ and $\Sigma(\not{p})$ are the self-energy diagrams given in 
sections~\ref{sec:ss}-\ref{sec:vv} and $m_{0,ij}^2$ is the tree-level mass squared.

There is an additional complication that arises because the Green functions 
and renormalization conditions are defined for the mass eigenstates. This means
we must find the mass eigenstates of the tree-level action first. We then renormalize
the mass eigenstates by solving either eqn.~(\ref{eqn:dets}) or~(\ref{eqn:detf}). This
will lead to a new rotation matrix which rotates the tree-level mass eigenstates into
the renormalized mass eigenstates. The mixing angles between the interaction eigenstates
and the renormalized mass eigenstates are then given by two rotations; the 
rotation from the interaction states to the tree-level mass states followed by a 
rotation from the tree-level mass states to the renormalized mass state.

\subsection{Default One-Loop Mass Corrections}

{\tmname{Effective}} provides two classes, \tmtexttt{Mass} and
\tmtexttt{MixingAngle}, which provide algebraic objects that can be used and
manipulated in expressions. Both of these classes require an object which
instructs them what level of approximation to compute the values at. By
default, if nothing is provided, the correction is given by the
\tmtexttt{MassCorrection::TreeLevel} class. As expected this class will simply
diagonalize the tree level mass matrix and return the value desired (mass or
mixing angle). We now discuss the three different one-loop corrections that
have been implemented in {\tmname{Effective}}. Again, a user can implement
their own approximation and renormalization conditions. Discussion of this 
extension is given in section~\ref{sec:custom}.

{\tmname{Effective}} provides three methods to generate one-loop mass
corrections. All of the corrections are an implementation of the on-shell
conditions. If a user wishes to implement alternate conditions, this can be
done by implementing their own \tmtexttt{MassCorrections} sub-class. Two of
the provided classes return the same one-loop mass, but different mixing angles.

We begin with the approximate method
\tmtexttt{OneLoopMassApprox}. This class approximates the
one-loop masses and provides only tree level mixing angles. This is done by
computing only the Green functions for the diagonal terms, i.e.
$\Pi_{i i}(p^2)$. This is computed with the tree level mass squared as
the argument for the momentum squared. The mass squared is then
\begin{equation}
  m^2_{i} = m_{0,i}^2 - \Pi(m^2_{0,i})_i, 
\end{equation}
where $\Pi(p^2)$ are the self-energy diagrams given in sections~\ref{sec:ss}-\ref{sec:vv}
and $m_{0,i}^2$ is the tree-level mass squared. For fermions this reads
\begin{equation}
m_i = m_{0,i} - \Sigma(m_{0,i})_i.
\end{equation}

The other methods solve eqn.~(\ref{eqn:dets}) or (\ref{eqn:detf}) using the bisection
method~\cite{NumericalRecipes}. The \tmtexttt{MassCorrection\-::OneLoopMass}
class returns the one-loop mass but only gives the tree-level mixing angles.
The \tmtexttt{MassCorrection::OneLoop} class returns both the one-loop mass and
mixing angle.

The implementation of the bisection method has five parameters which can be adjusted
by the user to improve numerical stability and convergence. The method first
brackets the solution. As an initial guess the starting point for the bracketing
is given by the tree-level mass plus-or-minus 1\%. In the case that the tree-level
mass is zero, the bracket is given by $\pm 10^{-9}$. Also, for efficiency, as many zero
mass fields are massless by construction, a quick check is performed. If the determinant
from eqn.~(\ref{eqn:dets}) or (\ref{eqn:detf}) is less than 
\tmtexttt{Numerics\-::MassRoundError} it is assumed that this mass is exactly zero. 
The default value of \tmtexttt{MassRoundError} is $10^{-4}$. 

If the mass is not zero and the initial bracket does not bracket the solution, than
the bracket is expanded until the solution is contained. If the expansion reaches
the maximum number of iterations, \tmtexttt{Numerics::MassMaxIterations}, than 
the method fails and the tree-level mass is returned and a failure message is printed.
Each iteration the bracket grows by $F \Delta x$ where $\Delta x$ is the current size of
the bracket and $F$ is the parameter \tmtexttt{Numerics::MassFactor}. The default
value is $0.6$. 

Once the solution is bracketed, the bisection method is iterated until the solution is
found, within \tmtexttt{Numerics::MassAccuracy} times the tree-level mass (if mass is 
zero then the accuracy is $10^{-9}$). The default value is $10^{-4}$, which means 
the solution is found within $0.01$\% of the value of the tree-level mass. If the 
method iterates \tmtexttt{Numerics\-::MassScanIterations} times and no solution is 
found, the method prints a failure message and returns the last value used.

\section{Renormalization Group Equations\label{sec:rge}}

In this initial release, {\tmname{Effective}} does not automatically generate
the renormalization group equations (RGE). The library does, however, provide
a class to handle RGEs and evolve the parameters according to those equations.
It is planned that in a future release, the one-loop RGEs will be
automatically generated for an arbitrary model according to one or more 
renormalization schemes.

This section will detail how one implements the RGEs of a model. The user
is free to implement these according to any renormalization procedure. It must
be noted that the default behaviour of the one-loop mass corrections must be
taken into account when implementing these RGEs. If one wishes to use a
different renormalization scheme, it may also be necessary to implement a new 
set of one-loop mass corrections in order to be consistent.

\subsection{Defining the Equations}

The class that handles the RGEs is the class \tmtexttt{RGE}. This class is
created by the constructor \tmtexttt{RGE(Model*)}; this constructor ties the
instance of the \tmtexttt{\tmtexttt{RGE}} class to a particular model. Upon
creation, the \tmtexttt{RGE} class creates a list of the parameters in the
\tmtexttt{Model} and associates with each parameter an expression.

The user can supply the expression for the $\beta$ function of any of the 
parameters to any order desired. This can be provided by a call to the routine
\tmtexttt{RGE::setBeta()}. This routine takes a \tmtexttt{Parameter} $x$ and
the expression for $\beta_x$. This routine saves the expression, in analytic
form, so that it can evolve the renormalization scale and adjust the parameter
$x$ appropriately. An example of defining and setting the appropriate $\beta$
function will be given now.

If we consider the one-loop $\beta$ function for the gaugino mass coupling
$M_1$ in the MSSM model. This is given by
\begin{eqnarray}
  \beta_{M_1} & = & \frac{11 g'^2 M_1}{8 \pi^2}, 
\end{eqnarray}
where $g'$ is the $U (1)$ coupling. This can be added to an \tmtexttt{RGE} object
by
\begin{verbatim}
RGE rge(&ew);
ex g1 = ew.getGaugeGroup("U1")->coupling();
Parameter M1 = ew.getParam("M1");
rge->setBeta(M1,11*g1*g1*M1/(8*Pi*Pi));
\end{verbatim}
We can now see how easy it is to implement the RGEs of a model, once they have
been calculated. This procedure can be used to redefine the gauge coupling
$\beta$ functions as well as all other couplings of the theory. Once the full
set of $\beta$ functions have been defined, we can then evolve between
different scales.

\subsection{Example of automatic RGE generation}
The currect version of \tmname{Effective} includes a prototypical example
of automatic RGE generation. When the object is created the $\beta$ functions of 
the gauge couplings are created automatically at one-loop. The one-loop $\beta$ 
function is
\begin{eqnarray}
  \beta_g^{1 - \tmop{loop}} & = & - \frac{g^3}{16 \pi^2} \left( \frac{11}{3}
  C_A - \sum_i \frac{4}{3} c_{i, g}^2 C_{F_i / A_i} - \sum_j \frac{1}{3} c_{j,
  g}^2 C_{F_j / A_j} \right) .
\end{eqnarray}
Here the sum over $i$ is the sum over fermions. $c_{i, g}$ is the charge of
fermion $i$ in group $g$. The factor $C_{F_i / A_i}$ represents the fact that
the fermion could be a gaugino (in supersymmetry) or a regular fermion. If it
is a gaugino it is in the adjoint representation and thus a factor of $C_A$,
otherwise the factor is $C_F$. The sum on $j$ is over all scalar fields with
the same factors as in the fermionic case.

It is important to note that this equation should also include $\theta (\mu -
m_i)$ where $\mu$ is the current renormalization scale and $m_i$ is the mass
of particle $i$. This would correctly implement the mass effects in the
$\beta$ function. The other couplings of a model do not yet have a $\beta$ 
function generated automatically. A future improvement is to include the mass
effects into the gauge couplings and to provide the full one-loop RGE for
an arbitrary model.

\subsection{Evolving Between Scales}

One of the powerful uses of the RGEs is to be able to define the parameters
at one renormalization scale, but use them to calculate at another. This is
especially important in SUSY models where one wants to define boundary
conditions at a high scale, i.e. $M_{\tmop{Planck}}$. These boundary
conditions are usually used to unify several parameters to one value at the
high scale, thus reducing the size of the independent parameter set.

In {\tmname{Effective}} the RGEs can be used to evolve the parameters between
different scales. The current scale of the model is given by a
\tmtexttt{Parameter} which is labeled as \tmtexttt{"renormScale"}. This scale
can be changed to a different scale and the parameters of the model are then
evolved to that scale by the RGEs. This is achieved by the routine
\tmtexttt{RGE::evolve()}. This takes a numeric value of the new scale and uses
the Runge-Carp-Kutta numerical method~\cite{NumericalRecipes} to iteratively apply the
differential equations of the RGE to the parameters so they are properly
evolved to the new scale.

The class \tmtexttt{RGE} also provides several routines which allow the user
to simply define the desired values for the parameters at a particular scale.
Then when the parameters are evolved to that scale, a simple call will force
all the parameters to take the preset values. This is very useful when
different parameters are defined at different scales and a consistent result
is desired.

For example, consider parameters $a$ and $b$. The $\beta$ functions of these
parameters are function of both $a$ and $b$. If we know $a$ should have a
specific value $a_0$ at scale $\mu_1$ and $b$ should have the value $b_0$ at
scale $\mu_2$, we must determine what $a$ is at $\mu_2$ and what $b$ is at
$\mu_1$ by an iterative approach. We must make some educated guess for $b
(\mu_1)$ and evolve to $\mu_2$. Our guess will be wrong so our $b (\mu_2)$ is
not equal to $b_0$. If we set it equal to $b_0$ and evolve to $\mu_1$ chances
are our $a (\mu_1)$ is not equal to $a_0$. We then set $a$ to $a_0$ and this
procedure can be iterated until the solutions converge, or we decide they are
not converging. If they don't converge we must assume that the boundary
conditions cannot simultaneously by fulfilled.

Such a procedure can be implemented in {\tmname{Effective}}. We will not refer
to a specific model in the next code block, but instead show how the routines
\tmtexttt{RGE::initialCondition()}, \tmtexttt{RGE::applyInitial()} and
\tmtexttt{RGE::evolve()} can be used to implement the above example.
\begin{verbatim}
double delta = 0.1;
double mu1 = 90.;
double mu2 = 120.;
double a0 = 5.;
double b0 = 7.;
Parameter a = 'model'->getParameter("a");
Parameter b = 'model'->getParameter("b");
rge.initialCondition(a,5.,mu1);
rge.initialCondition(b,7.,mu2);
bool consistent = false;
int max_tries = 50;
int tries = 0;
'model'->getParameter("renormScale")=mu1;
b = 1.;
while(!consistent && tries < max_tries) {
    rge->evolve(mu2);
    if((b+delta) > b0 && (b-delta) < b0) consistent = true;
    rge->applyInitial(mu2,3.);
    rge->evolve(mu1);
    rge->applyInitial(mu1,3.);
    tries++;
}
\end{verbatim}
We can see from this code that the routine \tmtexttt{RGE::initialCondition()}
allows us to set the value of a parameter at a scale. These initial conditions
can then be applied later by a call to \tmtexttt{RGE::applyInitial()}. We see
that this routine takes a scale as an argument. All \tmtexttt{Parameter}s that
were given an initial value are checked. If the scale that the initial
conditions were defined at is within $\delta$ of the given scale, the initial
value is applied. $\delta$ is the second argument to
\tmtexttt{RGE::applyInitial()}. If this is not provided the default value is
3.

Notice in this code that we have not designated any particular form to the
$\beta_a$ and $\beta_b$ functions. They must be defined for the code to work,
but their form does not impact the algorithm above. It only dictates if
consistent solutions can be found.

It is also important to note that the $\beta$ functions are stored
analytically. They can be printed in LaTeX format or in plain text by calling
the routine \tmtexttt{RGE::print()}. This routine takes a stream and prints
the $\beta$ functions to it. This can be useful when debugging your
definitions of the $\beta$ functions.

\section{The examination of arbitrary observables using \tmname{Effective}}


We have now discussed how to define the particle content of a model. We have 
enumerated the terms that this definition automatically supplies to the 
Lagrangian and explained how to include additional interaction terms and 
vacuum-expectation-values. We have seen how the effective potential and 
tadpoles can be accessed for this model both at tree-level and one-loop 
and how the potential can be minimized, and the tadpoles set to zero by 
numerical means. We have shown how the masses and mixing angles of the 
fields can be accessed and how the one-loop corrections are defined. Lastly, we 
have discussed the renormalization group equations and how they can be defined 
and used to give a consistent set of values at all scales.

The purpose of this library is to provide the user with a tool to study many 
different aspects of their model, yet we have not explained how to use 
effective to calculate physical observables other than the masses and mixing 
angles. \tmname{Effective} provides an abstract class \tmtexttt{Diagram} that
can be used to compute physical observables deriving couplings and parameters 
from the model.  The \tmtexttt{Diagram} class is not the only way that physically
relevant information can be drawn out of \tmname{Effective}. A creative user may 
be able to find interesting and exciting uses for this library well beyond the 
scope of this article or the authors' imaginations!

As mentioned, the feature built into {\tmname{Effective}} which makes it easy to study 
arbitrary properties of a a model is the \tmtexttt{Diagram} class. The user 
specifies the expression corresponding to some diagram, and then effective 
automatically iterates through all appropriate fields on internal lines, and all 
appropriate couplings (with the right spin and derivative structure) at each vertex. 
The class is used internally by {\tmname{Effective}} with the Passarino-Veltman 
functions to obtain masses and mixing angles at one-loop, so the user can use 
these as examples when creating their own diagrams.

The class is an abstract class; some of the routines of the class must be filled in 
by the user in a class which is derived from \tmtexttt{Diagram}. The main interface 
to the class is through the method \tmtexttt{evaluate()}. This routine will call the 
abstracted routines appropriately, building up a summed value for the diagram as it 
loops through all $n$ point couplings and fields on internal lines in a model by 
calling the abstract function \tmtexttt{calculate()} for each coupling to determine its value. 

The routines that the user must define in the deriving class are 
\tmtexttt{calculate()}, \tmtexttt{function()}. 
The \tmtexttt{calculate()} routine is called by the \tmtexttt{evaluate()} routine. 
On input, an $n$ point coupling is provided, the routine should check that the 
coupling is appropriate to the diagram being defined. It is the responsibility of 
the class which inherits \tmtexttt{Diagram} to specify which couplings are appropriate 
to the diagram. For example in the \tmtexttt{FourPtLoop} class the coupling must 
correspond to the desired two external fields and have the same internal field. 

In the current version of \tmname{Effective} the sums over fields on internal propagators 
must be carried explicitly in the user-written \tmtexttt{calculate()} routine; those 
implementing new diagrams should copy over the appropriate loop statements. In a later 
version we hope to refactor this into the automatic parts of the \tmtexttt{Diagram} base 
class. 

The \tmtexttt{Diagram} class provides a few routines which can be helpful when determining 
if the given coupling matches the desired criteria. The routine 
\tmtexttt{Diagram\-::makeList()} will take the expression for the fields of a coupling 
and return a list; each element contains a flag indicating if the field was accompanied 
by a derivative, and the field itself. The routine \tmtexttt{Diagram::find()} can be 
used to determine if a field with a particular derivative flag is in the list returned 
by \tmtexttt{makeList()}. If it is, the routine \tmtexttt{Diagram::remove()} can be 
used to remove the matching item. When the routine has decided whether the coupling 
is appropriate, it should call \tmtexttt{function()}.

\tmtexttt{function()} is intended to be the value of the diagram for a
particular choice of vertex and spins of internal fields. This idea can be seen in the mass
corrections of section \ref{sec:massm}. In those diagrams, the function is
defined for a particular choice of the spin of the intermediate particles.
Then this function depends only on the $p^2$ of the Green's function and the
masses of the intermediate fields.

The best way to understand how the \tmtexttt{Diagram} class is intended to be
used is by a demonstration. This is too long to include in this article, and is
instead deferred to a tutorial page on the website mentioned in the first
section.

\section{Practicalities\label{sec:uses}}

In this section we will discuss some practical issues with Effective.
We will describe routines which reduce processor overhead by saving and 
loading the system's internal representation of a model: the coupling database 
and the parameter values. We will also discuss the rudamentery command line 
interface and how to achieve the same effects without relying on the command 
line interface.

Before we discuss these uses we must discuss one essential routine, 
\tmtexttt{Model\-::initialize()}. This routine {\tmem{must}} be run before any
physics is computed. This routine takes no arguments and specifies to the
model that all the information needed to generate the Lagrangian and mass
matrices is present. Once this is run the user may proceed to study their
model as they wish.

Of course the purpose of this library is to provide the user with a tool to
study many different aspects of their model. To this end, the discussions in
this section and the next do not contain all possible uses. A creative user
may be able to find interesting and exciting uses for this library well beyond
the scope of this article.

\subsection{Saving and Loading Couplings}

\tmname{Effective} generates internally a coupling database from the action for
use in calculating observables and mixing angles. This can grow to be quite large.
For large models the generation of this database can take of order of 30 minutes on a modern
computer. The results of this generation can be saved to a file, however, and
on future runs this can be read in just a few seconds. We discuss here how to
specify whether to generate the couplings or read them. We also discuss when
the database {\tmem{must}} be generated.

The \tmtexttt{Model} class has a variable
\tmtexttt{Model::couplingsFromScratch} which is a \tmtexttt{static} member of
the class . If this is set to \tmtexttt{true} before the model is initialized,
then the couplings will be generated directly from the Lagrangian. If this is
\tmtexttt{false} then the couplings will attempt to be loaded from a file. The
default value of this variable is \tmtexttt{false}.

If one wishes to load the couplings from a file, the filename must be
specified before initializing the model. This is done by calling the
\tmtexttt{Model::couplings()} routine. This takes the filename as an argument.
The same routine is used to specify a filename to save a file to. If the
\tmtexttt{static} variable \tmtexttt{Model::saveCouplings} is true, then the
model will automatically save the file when the program completes. These
features are illustrated by the following code.
\begin{verbatim}
ElectroWeak ew;
Model::couplingsFromScratch = true;
Model::saveCouplings = true;
ew.couplings("ew.gar");
ew.initialize();
\end{verbatim}
These features are also accessible through the command line interface. A
simple routine \tmtexttt{Model::readCommandLine()} has been written that
accepts the flags \tmtexttt{-r}, \tmtexttt{-s} and \tmtexttt{-h}. The
\tmtexttt{-r} flag is used to specify that the couplings should be
regenerated. The \tmtexttt{-s} flag specifies that the couplings should be
saved to the file and the \tmtexttt{-h} flag is a help command which prints
all options. This is a very simple interface but it allows the user to run the
code with different behaviours without having to recompile.

If we rewrite the above code as
\begin{verbatim}
ElectroWeak ew;
Model::readCommandLine(argc,argv,&ew);
ew.couplings("ew.gar");
ew.initialize();
\end{verbatim}
we can now run the executable in several different variants. If we wish to
load the couplings, we provide no flags. If we want to save the couplings we
give the \tmtexttt{-s} flag and is we want to recompute the couplings, we give
the \tmtexttt{-r} flag. It is obvious that most times, the \tmtexttt{-r} and
\tmtexttt{-s} flags will accompany each other.

It is also important for the user to know when they should recompute the
couplings. The couplings must be recomputed if the Lagrangian is modified
{\tmem{at all{\tmem{}}}}. For example, one may wish to begin studying a model
with only one family of fermions. They may wish to then include 3 families of
fermions and repeat their studies. Once they include the additional terms in
the Lagrangian they must recompute the couplings. Failure to do this will mean
that they will only load the couplings for their one family model and their
studies will be wrong.

\subsection{Modifying, Saving and Loading Parameters\label{sec:params}}

One of the useful features of {\tmname{Effective}} is the ability to keep the
expressions in analytic form and to perform numeric operations on these
expressions. This is able to be done because all \tmtexttt{Parameter}s are
simultaneously an algebraic object and a numeric value. The object is kept in
all expressions until the \tmtexttt{evalf()} function is called. At that
point, the algebraic object is replaced by the numeric value of the parameter.

This duplicity of the \tmtexttt{Parameter} object allows the user to create
all their expressions, derived from the Lagrangian, and very quickly access
the numeric value of the expression. Changing the value of a
\tmtexttt{Parameter} globally changes the value all expressions. When
\tmtexttt{evalf()}'d an expression will return its numeric value with the changed value
for the parameter. For example, changing the value of a VEV will
simultaneously change the value of the effective potential and the masses of
the particles.

It is important to note that though the undiagonalized mass matrix is
automatically changed to reflect the new parameter value, the
{\tmem{diagonalized}} mass matrix may not be. If the mass matrix is larger
than $2 \times 2$, then the diagonalization is a numeric routine. This means
it must be recomputed to update the mass of a particle. Not to worry though, a
simple call to the \tmtexttt{Model::resetMasses()} routine will instruct the
library to rediagonalize all matrices which are diagonalized numerically.

To illustrate the power of this feature, we will give an example of how to
evaluate the one-loop effective potential as a function of the VEV.
Considering again the electro-weak model the following code scans the VEV and
evaluates the one-loop effective potential.
\begin{verbatim}
Parameter vev = ew.getParam("HiggsVev");
double v;
for(v = -246.; v<=246.; v+=5.) {
   vev = v;
   ew.resetMasses();
   cout << v << "\t"
        << ew.potential(Approximation::OneLoop).evalf()
        << endl;
}
\end{verbatim}
We see in this code that to set the value of a \tmtexttt{Parameter}, we treat
it like we would a normal variable. A simple call to the \tmtexttt{=} operator
sets the numeric value of the VEV. This is then propagated to all the masses
(used to compute the one-loop potential) and the tree-level potential
expression. Simply calling \tmtexttt{evalf()} on the potential replaces all
the \tmtexttt{Parameter}s with their numeric value.

There is also a simple input/output mechanism for the values of the
\tmtexttt{Para\-meter}s of a model. This saves or loads the values from a tab
delimited file. This is achieved by the routines
\tmtexttt{saveParameters()} or \tmtexttt{loadParameters()} of the
class \tmtexttt{Model}.
These routines take the stream to read or write to. The
\tmtexttt{loadParameters()} routine will print an error message if the file
does not have the correct structure or one of the \tmtexttt{Parameter}s has
the wrong label. We give an example of the one-family electro-weak input file.
\begin{verbatim}
renormScale   91.0
HiggsVev      246.0
SU2           0.653089
U1            0.3550
Ye            1.3e-5
lambda        0.182513
mu            105.095
\end{verbatim}
We notice that the first line of the file is the renormalization scale. This is
also the scale for all of the parameters in this file. We saw in section
\ref{sec:rge} a possible treatment for parameters defined at different scales,
these simple input/output routines are insufficient for input at multiple scales.

It is possible to save the parameters in any format desired, for example
according to the SUSY Les Houches Accord~\cite{Skands:2003cj} (LHA). This simply requires 
writing new code to print and read the format appropriately. The Les Houches Accord format
has not been implemented in this version of {\tmname{Effective}}, but the SUSY 
LHA format is a planned upgrade.

\section{Example: One Family Electro-Weak Model\label{sec:ew}}

We now turn to a concrete example, in its entirety. Throughout this manual we
have defined pieces of the one-family electro-weak model. The full listing of
the class definition can be found in the appendix. Here we present the main
code block where the physics analysis begins. We will then give two examples
of using the model to study something of physical relevance. These examples
are not intended to be useful physics studies, only illustrative examples of
using a model.

\subsection{Main Code}

We have defined the model \tmtexttt{ElectroWeak} in the file \tmtexttt{EW.h},
found in the appendix. We now give the code which initializes the model and
allows us to begin a physics analysis.
\begin{verbatim}
#include "EW.h"
#include <effective/effective.h>

int main(int argc, char **argv) {
   try {
      // Begin by initializing the model
      ElectroWeak ew;
      Model::readCommandLine(argc,argv,&ew);
      ew.couplings("ew.gar");
      ew.initialize();

      // Now lets load the values of the parameters
      ifstream parin;
      parin.open("EW.dat");
      ew.loadParameters(parin);
      parin.close();

      // Lets create a LaTeX stream, ew.tex
      ofstream fout;
      Utils::startLatex("ew.tex",fout);

      // ... insert physics analysis code here ...

      // now lets clean up
      Utils::closeLatex(fout);
   } catch(exception &p) {
      cerr << p.what() << endl;
      return 1;
   }
   return 0;
}
\end{verbatim}
We will now discuss the new items in this code. We see that the file
\tmtexttt{effective.h} must be included in order to use {\tmname{Effective}}.
We also see that two new routines \tmtexttt{Utils::startLatex()} and
\tmtexttt{Utils::closeLatex()} have been used. The \tmtexttt{startLatex()}
routine takes a filename and a stream. It opens the file into the stream and
sets the behaviour so that all expressions will be printed in LaTeX format.
One must then remember to set and unset the math mode of LaTeX appropriately,
depending on what is being saved to the file. This routine also prints a
preamble so that the LaTeX file can be used without editing to produce a
document. The \tmtexttt{closeLatex()} then prints the
\tmtexttt{\tmbsl end\{document\}} string and closes the stream.

The code also contains the \tmtexttt{try \{ ... \} catch \{ ... \} } statements in the
code. {\tmname{Effective}} uses the C++ exception handling mechanism to deal
with unusual behaviour and internal errors. The try/catch clause catches any
errors that have not been dealt with. The catch block then prints the error
message, hopefully allowing the user to understand what failed. If these
try/catch clauses are not included, the errors may just cause a crash without
any information why.

\subsection{Case Study 1: Higgs Potential vs. $\upsilon$}

\begin{figure}[h]
  \epsfig{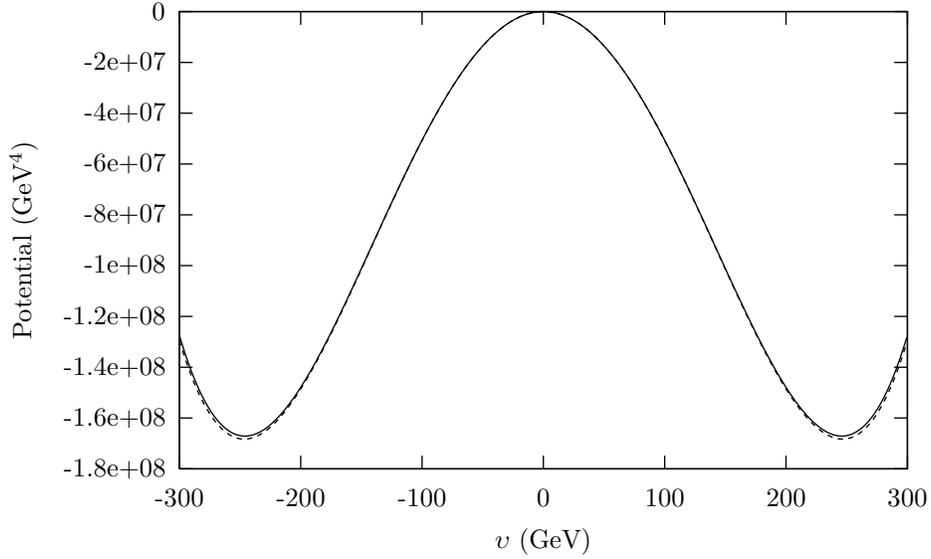}
  \caption{The dependence on the tree-level (solid) and one-loop (dashed) effective 
  potential as a function of $\upsilon$.\label{fig:higgspot}}
\end{figure}

We now turn our attention to the first of two examples of a physics analysis.
This first one will produce a tab-delimited file which can be used with a
plotting program to give the tree-level and one-loop potential as a function
of the VEV. This is given by
\begin{verbatim}
Parameter vev = ew.getParam("HiggsVev");
ofstream datafile;
datafile.open("potData.dat");
for(double v = -246.0; v<=246.0; v+=0.5) {
   vev = v;
   ex tree = ew.potential(Approximation::TreeLevel);
   ex one = ew.potential(Approximation::OneLoop);
   datafile << v << "\t" << tree.evalf() << "\t" << one.evalf()
            << endl;
   ew.resetMasses();
}
datafile.close();
\end{verbatim}
Using the values of the parameters given in section \ref{sec:params} this code
was used to produce figure \ref{fig:higgspot}.

\subsection{Case Study 2: $\cos \theta_W$ as a function of the $S U (2)$
coupling}

\begin{figure}[h]
  \epsfig{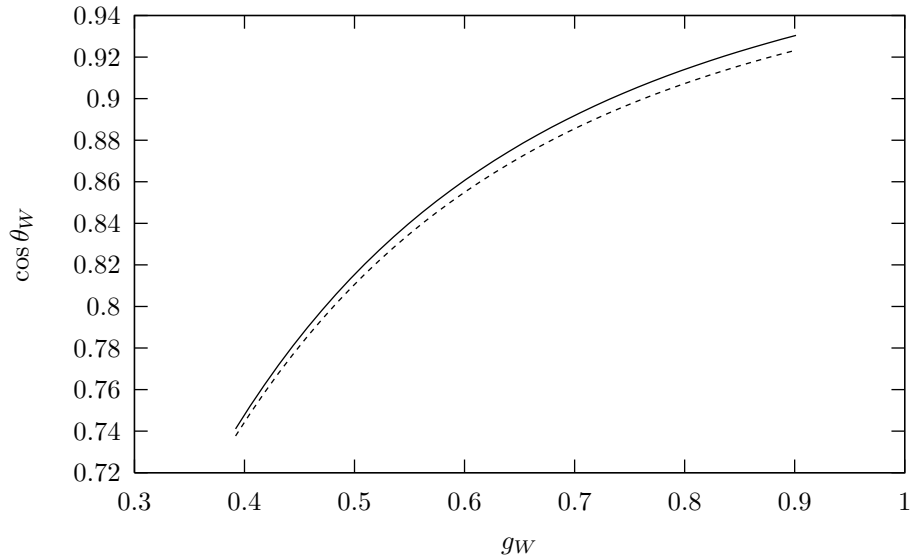}
  \caption{Plot of $\cos \theta_W$ vs the $S U (2)$ coupling parameter for the
  tree-level masses (solid) and the one-loop masses (dashed). All
  values correspond to the renormalization scale $\mu = 91.2
  \tmop{GeV}$.\label{fig:sinw}}
\end{figure}

We now turn to another observable. In this case we will look into the Weinberg
angle, $\theta_W$. The definition of the angle is $M_W = M_Z \cos \theta_W$.
Thus we can study the ratio $M_W / M_Z$ at tree-level or one-loop. The
following code produces a tab-delimited file which can be plotted. The
result is shown in figure \ref{fig:sinw}. In this example we fix the $U (1)$
coupling to 0.355 and the Higgs VEV, $\upsilon$, to 246.0 GeV. All values are for
the renormalization scale $\mu = 91.2 \tmop{GeV}$.
\begin{verbatim}
ofstream cosW;
cosW.open("cosW.dat");
double lowg, highg;
double fourtyPct = ew.getParameter("SU2").to_double()*0.4;
lowg = ew.getParameter("SU2").to_double()-fourtyPct;
highg = ew.getParameter("SU2").to_double()+fourtyPct;
idx i = ew.getGaugeGroup("SU2")->gaugeIndex();
ex W = ew.getGaugeField("W")->expression().subs(i==2);
ex Z = ew.getGaugeField("W")->expression().subs(i==3);
ex MWt = Mass(&ew,W,MassCorrections::treeLevel);
ex MWo = Mass(&ew,W,MassCorrections::oneLoop);
ex MZt = Mass(&ew,Z,MassCorrections::treeLevel);
ex MZo = Mass(&ew,Z,MassCorrections::oneLoop);
for(double g = lowg; g<highg; g+=fourtyPct/20.) {
   ew.getParameter("SU2") = g;
   ew.resetMasses();
   cosW << "\t" << MWt.evalf()/MZt.evalf() << "\t" 
        << MWo.evalf()/MZo.evalf() << endl;
}
\end{verbatim}

\section{Customizability\label{sec:custom}}

Throughout this text we have referred to {\tmname{Effective}}'s modular design
and the ability for a user to create extensions to study many different
classes of models. One extension that may be of use is to implement a
different group structure. Implementation of such a class is given in Appendix
B.3 of~\cite{Stephens:2004mf} and repeated in tutorials on the website.

What we will discuss here is how to implement new objects to provide
behaviours that differ from the default implementations. This includes a new 
subclass of the \tmtexttt{Spin} class as well
as new \tmtexttt{Field} classes. One may wish to implement special operators
which act on terms in the Lagrangian. This is also discussed below. Lastly, a
user may wish to use their own renormalization scheme and thus provide a
different set of one-loop corrections. They may also have a closed form of a
mass correction to a given order. These types of customizations are discussed
here.

\subsection{New Operators}

{\tmname{Effective}} has a built in treatment of Lorentz derivatives. This
includes knowledge of how to treat the Lorentz indices in compound terms which
contain several Lorentz derivatives and vector fields. A user can define their
own operator, but if this operates on fields, then the appropriate treatment of
the operator must be included in the \tmtexttt{Spin} class as discussed below.

In this part we will simply discuss what is needed to define the actual
operator. We will take, as a concrete example, the idea of a light-cone
operator, $\left(n^\pm\right)^\mu$. This operator takes any Lorentz vector 
and projects it to either the
plus or minus part of the light cone. In this example, the light-cone operator
needs two arguments, the object which is being projected (e.g. momentum or
vector field) and the sign.

Code details of the definition of this object can be found in the
{\tmname{GiNaC}} tutorials~\cite{GiNaC}. Here we present only the code and how the
different parts cause the behaviour we desire. We begin with the class
definition.
\begin{verbatim}
const unsigned TINFO_LightCone = 0x1100008U;

class LightCone : public basic {
 GINAC_DECLARE_REGISTERED_CLASS(LightCone,basic);

private:
  bool itsSign;
  ex itsField;

public:
  LightCone(const ex &arg, bool plus);
  LightCone(const LightCone &lc);

  // virtual functions 
  void do_print(const print_context &c, unsigned level = 0) 
        const; 
  void do_print_latex(const print_context &c, 
        unsigned level = 0) const; 
  ex eval(int level=0) const; 
  ex evalf(int level=0) const; 
  ex op(size_t i) const; 
  ex & let_op(size_t i) { return itsField; } 
  size_t nops() const;
  bool sign() const { return itsSign; } 
};
\end{verbatim}
We can see that there are only a few functions that need to be implemented.
Here we give the implementation of the constructor and the comparison
operator. The other routines are trivial.
\begin{verbatim}
GINAC_IMPLEMENT_REGISTERED_CLASS_OPT(LightCone,indexed,
 print_func<print_context>(&LightCone::do_print).
 print_func<print_latex>(&LightCone::do_print_latex));

LightCone::LightCone(const ex &arg, bool p) 
 : basic(TINFO_LightCone), itsSign(p) { 
  // Simply replace the lorentz index by a + or - 
  symbol lst idxs; 
  LorentzStructure::getIndices(arg,idxs); 
  // Indicate an error for non Lorentz vector objects
  if(idxs.size() != 1) {
    cerr << "Called a light cone operator on a "
         << "Lorentz tensor with rank != 1.\n";
  }
  itsField = arg.subs(idxs[0]==LorentzStructure::lorentzIndex(0));
}

int LightCone::compare_same_type(const basic &other) const {
  int signdiff; 
  const LightCone &o = static_cast<const LightCone &>(other);
  if(o.sign() == sign() signdiff = 0;
  else signdiff = o.sign() ? 1 : -1;
  int valdiff = itsField.compare(o.itsField);
  if(valdiff) return valdiff;
  else return signdiff;
}
\end{verbatim}
Here we see that in order to get a functional operator we have simply had to
properly define the comparison of two objects (in combination with the
routines in the header file, which are trivial). These few simple routines
provide an operator which acts on an expression from {\tmname{GiNaC}} and is
treated as its own algebraic object in the engine.

Of course, this particular implementation may not be wildly useful. Instead
one may which to write, in the \tmtexttt{eval()} routine, code which applies
to distributive property to the expression passed to it. So for example
$(n^+)^{\mu} \left( A_{\mu} + B_{\mu}) \right.$ properly returns $A^+ + B^+$.
Again these issues are explained in detail in the {\tmname{GiNaC}} manual and
tutorials.

\subsection{New Spin Classes}

As discussed in section \ref{sec:field}, the \tmtexttt{Spin} class determines
several properties which determine what terms enter the Lagrangian. It also
provides routines which are used to derive the couplings of fields. In order
to change these behaviours, a user must implement a new subclass of
\tmtexttt{Spin}. Here we will discuss the \tmtexttt{Spin} class in detail and
explain what each routine is used for. This will be a valuable reference for a
user who wishes to implement new types of fields.

We begin with table \ref{tab:spin}. Here we see the full list of routines in
the \tmtexttt{Spin} class and whether each routine is \tmtexttt{virtual} or
not. The \tmtexttt{virtual} routines can be reimplemented by a subclass of
\tmtexttt{Spin}.

\begin{table}[h]
  \tmtexttt{}\begin{tabular}{llll}
    & {\tmstrong{Routine}} & {\tmstrong{Arguments}} & {\tmstrong{Return}}\\\hline
    \tmtexttt{virtual} & \tmtexttt{interaction} & \tmtexttt{Field*} &
    \tmtexttt{ex}\\
    \tmtexttt{virtual} & \tmtexttt{isLorentzIndex} &  & \tmtexttt{bool}\\
    \tmtexttt{virtual} & \tmtexttt{imagPart} &  & \tmtexttt{bool}\\
    \tmtexttt{virtual} & \tmtexttt{massCorrection} & \tmtexttt{ex, ex, ex,
    Model*}, & \tmtexttt{ex}\\
    &  & \tmtexttt{bool, bool} & \\
    \tmtexttt{virtual} & \tmtexttt{lorentzDerivativeCouplings} & \tmtexttt{ex,
    Couplings*, ex,} & \tmtexttt{epair}\\
    &  & \tmtexttt{ex, int, int} & \\
    \tmtexttt{virtual} & \tmtexttt{operator==} & \tmtexttt{const Spin \&} &
    \tmtexttt{bool}\\
    \tmtexttt{virtual} & \tmtexttt{clone} &  & \tmtexttt{Spin*}\\
    \tmtexttt{virtual} & \tmtexttt{alternateOperatorCouplings} & \tmtexttt{ex,
    Couplings*, ex,} & \tmtexttt{epairv}\\
    &  & \tmtexttt{ex, int} & \\
    \tmtexttt{virtual} & \tmtexttt{massMatrixCoeff} &  & \tmtexttt{ex}\\
    \tmtexttt{virtual} & \tmtexttt{isMassSqrt} &  & \tmtexttt{bool}\\
    & \tmtexttt{spin} &  & \tmtexttt{double}\\
    & \tmtexttt{coeff} & \tmtexttt{Couplings*, ex, ex,} & \tmtexttt{ex}\\
    &  & \tmtexttt{ex, ex\&, int, bool} & \\
    & \tmtexttt{getName} &  & \tmtexttt{string}
  \end{tabular}
  \caption{Complete list of routines for the \tmtexttt{Spin} class. Details of
  each routine can be found in the text.\label{tab:spin}}
\end{table}

We discuss here what each of the routines of the \tmtexttt{Spin} class means 
and how a new implementation can be made. We will begin with the
virtual routines which have a default behaviour. This means that these
routines do not need to be reimplemented by the user, unless they wish to
change their behaviour.
\begin{itemizedot}
  \item \tmtexttt{isLorentzIndex():} This routine indicates whether fields of
  this spin have a Lorentz index or not. The default value is
  \tmtexttt{false}.
  
  \item \tmtexttt{imagPart():} This specifies whether a field should have a
  real and imaginary component (\tmtexttt{true\tmtexttt{}}) or just a real
  component (\tmtexttt{false}). The default value is \tmtexttt{true}.
  
  \item \tmtexttt{alternateOperatorCouplings():} This routine is used to find
  the couplings of a field of this spin in the Lagrangian when non-default
  operators have been implemented. This means that if the user has implemented
  their own operator which acts on terms in the Lagrangian, they must also
  implement this routine for all field types which are operated on by the new
  operator. This routine takes arguments we will label here as \tmtexttt{f, c,
  x, idx, level}. The derivation of couplings is a recursive procedure in
  {\tmname{Effective}}. The procedure which derives the couplings will call
  this routine at every level of recursion. It is the responsibility of this
  routine to return a list of expression pairs (\tmtexttt{epairv}) which each
  contain a new index and value for the coupling. This coupling is the
  derivative of \tmtexttt{x} with respect to the new operator $O_i$ acting on
  \tmtexttt{f}, i.e. $\frac{\partial x}{\partial O_i (f)}$. Each element of the list
  is for each additional new operator that has been defined. The new index is
  given by the old index, \tmtexttt{idx}, times the field acted upon it by
  $O_i$. The \tmtexttt{level} argument is an integer which indicates how many
  levels of recursion are still to be done. In some cases this can be quickly
  used to decide whether to perform the derivative on \tmtexttt{x} or not,
  thus speeding up the routine considerably. By default this routine returns
  an empty list.
  
  \item \tmtexttt{massMatrixCoeff():} This routine returns an expression which
  is a coefficient used for the mass matrix calculation. The mass matrices are
  computed by eqn. (\ref{eqn:massm}), where the coefficient is simply the
  sign. This factor is usually just a sign, but this implementation allows it
  to be whatever the user requires. The default behaviour is just to return
  $1$.
  
  \item \tmtexttt{isMassSqrt():} This routine indicates whether the mass
  matrix requires a square root (\tmtexttt{true}) to return the mass or not.
  For example the vector and scalar matrices give the mass squared whereas
  the fermion mass matrix gives only the mass. The default value is \tmtexttt{true}.
\end{itemizedot}
We now turn to the non-virtual routines. These cannot be overwritten. Instead,
they are just simple wrappers or data access routines. The constructor for a
\tmtexttt{Spin} object takes a \tmtexttt{string} and a \tmtexttt{double}
value. This string is the name of the spin type and the double represents the
spin factor, i.e. $(2 s - 1) (- 1)^{2 s}$. The non-virtual routines are:
\begin{itemizedot}
  \item \tmtexttt{spin():} This routine returns the value of the spin factor
  given to the constructor.
  
  \item \tmtexttt{getName():} This routine returns the name of this spin type
  given to the constructor.
  
  \item \tmtexttt{coeff():} This routine is a wrapper which takes the
  \tmtexttt{Couplings} pointer as well as the field and expression to
  differentiate and performs the differentiation. This routine also takes an
  extra argument, the \tmtexttt{ex\&}, which is modified to represent the new
  index after differentiation. The Lorentz structure of the index may not be
  unique in a brute force type of implementation. In order for the index to be
  unique for a given set of fields the differentiation is performed and the
  appropriate transformations of the Lorentz indices is applied so that the
  index is always unique for a given set of fields, and the coupling times
  that unique index is the correct expression from the Lagrangian. The final
  argument is a boolean which indicates whether to let the
  \tmtexttt{Couplings} class handle the Lorentz index substitutions
  (\tmtexttt{false}) or not.
\end{itemizedot}
We finally now turn our attention to the virtual functions which have no
default implementation in the \tmtexttt{Spin} class. These routines
{\tmem{must}} be reimplemented by a \tmtexttt{Spin} subclass in order to be
used.
\begin{itemizedot}
  \item \tmtexttt{interaction():} This returns the kinetic terms of the field
  of this spin type. This is the routine that must be changed implement
  different terms in the Lagrangian.
  
  \item \tmtexttt{lorentzDerivativeCouplings():} This is similar to the
  previously discussed \tmtexttt{alternateOperatorCouplings()} routine. This
  routine is only for Lorentz derivatives and rather than return a list of
  expression pairs it returns only one (\tmtexttt{epair}).
  
  \item \tmtexttt{operator==():} This is a simple comparison routine. If the
  two spin objects are identical, then \tmtexttt{true} is returned.
  
  \item \tmtexttt{clone():} This routine creates a new object which is
  identical to the one it is called on. This routine dynamically allocates a
  new pointer, it is the responsibility of the calling method to handle the
  memory deallocation.
\end{itemizedot}
Now we have discussed in detail the methods of the \tmtexttt{Spin} object we
must explain how a new spin class can be included into the framework of
{\tmname{Effective.}} The default behaviour of {\tmname{Effective}} is to
include the \tmtexttt{ScalarSpin}, \tmtexttt{FermionSpin} and
\tmtexttt{VectorSpin} spin types. A user can add a new \tmtexttt{Spin}
subclass by calling \tmtexttt{Model::addSpin()} before the
\tmtexttt{initialize()} routine is called. A logical place is in the
constructor or the \tmtexttt{createMatterFields()} or
\tmtexttt{createGaugeFields()} routines. The \tmtexttt{addSpin} routine will add the new spin to
the list of spins. Each spin type in this list has a corresponding mass
matrix. This means that only fields with the same \tmtexttt{Spin} classes will
mix.

It is also possible to create a new \tmtexttt{Spin} class which replaces on of
the default ones. In order to tell {\tmname{Effective}} to use the new class
instead of the old one, the user must call \tmtexttt{Model::changeSpin()}.
This routine takes the new spin object and an index to one of the old ones.
This index is either \tmtexttt{Model::Scalar}, \tmtexttt{Model::Fermion} or
\tmtexttt{Model::Vector}. Again, this routine needs to be called before the
\tmtexttt{initialize()} routine.

\subsection{New Field Behaviours}

Another major modification that a user may want to implement it to define a
new type of field. For example, currently {\tmname{Effective}} does not
provide a way for a field to be in the adjoint representation of a group and
be charged under other groups. This is done, for example, in models with Higgs
triplets~\cite{Forshaw:2003kh}.

Here we will give the outline as to the method to create such a field. The
\tmtexttt{MatterField} will be our starting reference. In fact, this almost
completely describes our new field, except we need a list of flags to indicate
which groups are in the adjoint representation and which are in the
fundamental. One would then, when creating the list of indices, add the
adjoint index to the list (which is retrieved by
\tmtexttt{GaugeGroup::gaugeIndex}) rather than the fundamental index
(retrieved with \tmtexttt{GaugeGroup::matterIndex}). The only other changes
required would be to the \tmtexttt{covariantDerivative()} and
\tmtexttt{susy()}, where instead of using the generators, the structure
functions would need to be applied for the adjoint representation. Lastly, the
\tmtexttt{Field::Cr()} and \tmtexttt{Field::C2()} functions would have to be
defined to return the correct values for the adjoint representation.

As long as this new class is a subclass of \tmtexttt{Field} we can immediately
use it with {\tmname{Effective}} to study this new class of model. In fact,
such an extension will be included in a future version of {\tmname{Effective}}
along with improvements to the \tmtexttt{Field} class to allow for a more
natural extension to $N \neq 1$ SUSY and higher rank tensor fields. Though all
of these are possible now, it would be difficult to make everything work in a
natural way. Changing some underlying structures will simplify such
extensions.

\subsection{New One-Loop and Beyond Mass Corrections}

The last major customization a user can implement is the mass corrections. The
\tmtexttt{Mass} class constructor takes three arguments. The model, the field
and a \tmtexttt{MassCorrection} class. This class has one \tmtexttt{virtual}
routine, \tmtexttt{operator()}, that must be implemented. In
{\tmname{Effective}} there are four types of \tmtexttt{MassCorrection} classes
implemented. These are discussed in section \ref{sec:massm}. Here we discuss
what \tmtexttt{operator()} calculates so that users can implement their own
version.

The \tmtexttt{operator()} routine has the arguments:
\tmtexttt{BlockMatrix \&}, \tmtexttt{Spin*}, and \tmtexttt{Model*}.
This routine is called by the \tmtexttt{Mass} and \tmtexttt{MixingAngle}
classes when evaluated. The \tmtexttt{BlockMatrix} class contains the
undiagonalized, diagonalized and the rotation matrix for tree-level and
matrices for the correction (at any order) and the diagonalized result and the
rotation matrix (also at any order). It is the \tmtexttt{operator()} routines
job to take the undiagonalized tree-level matrix and generate the mixing
matrix and the diagonal values of the matrix, at any order. It is then up to
the user to implement their corrections and the diagonalization of the mass
matrix how they wish.

This procedure will become clear after we look at the \tmtexttt{OneLoopMassApprox}
implementation. This class gives the approximate one-loop mass, but only the
tree-level mixing matrix. This is done by computing only the diagonal
corrections to the diagonalized mass matrix.
\begin{verbatim}
void OneLoopMass::operator()(BlockMatrix &bm, Spin *s, 
                             Model *m) {
   if(MassMatrix::rediag(bm)) {
      int temp;
      matrix r = bm.undiag;
      Utils::diagonalize(r,bm.diag,bm.rotate,temp);
   }
   if(bm.lastFlag != flag) {
      Utils::matrixCopy(bm.correction,correction(bm,m,s));
      Utils::matrixCopy(bm.crotate,bm.rotate);
      Utils::matrixCopy(bm.cdiag,bm.correction);
      ftime(&bm.lastDiag);
      bm.lastFlag = flag;
   }
}
\end{verbatim}
This code simply checks that the parameters haven't changed, and if they
have rediagonalizes the tree level matrix. If the last evaluation wasn't the
same approximation as this one, this method then computes the corrections to
matrix. From this one can imagine how they may
implement a different type of correction. For example, the
\tmtexttt{correction()} routine which is called to fill the correction matrix
with its values, may contain clauses to identify specific particles and apply
corrections derived from a paper for a particular particle.

\section{Future Outlook}

This article has been designed to serve a few purposes. We have described many
of the important tools that allow a user to probe the physics of their model as
they see fit. This gives users a manual by which to begin to implement their
model.

We have also explained the assumptions and physics behind the code. This way
a user is able to know what to expect from the default behaviour of the
library, and design extensions that suit their needs. We have discussed
several extensions that are possible. These are the ones that we feel are
likely to be most useful for users.

We have not provided detailed code documentation nor a list of all possible
extensions. This has been reserved for the website, which will contain
tutorials with detailed code.

Throughout this text we have discussed future improvements we plan to make on
the code. The status of such improvements can be found on the webpage. The
most pressing upgrades we would like to implement are the following:
\begin{itemize}
  \item We plan to generalize the \tmtexttt{Field} implementation so a wider
  range of models can be easily implemented.
  
  \item For many models some symmetries, for example $S U (3)_C$, are
  unbroken. It may be desirable to not explicitly sum over the indices of
  these groups, except when computing numerical results.
  
  \item Properly provide a supermultiplet object which can be used to define
  the superpotential.
  
  \item Automate the one-loop RGE for at least one renormalization scheme.
\end{itemize}

\vspace{10mm}
\noindent
{\bf Acknowledgments}\vspace{2mm}\\
JH acknowledges the support of PPARC studentship
number PPA/S/S/1999/02798.
 
\appendix

\section{\tmtexttt{ElectroWeak} definition}

We present here the full definition of the \tmtexttt{ElectroWeak} class. Here
we define it in one file, \tmtexttt{EW.h}, though the class definition and the
routine implementations could be separated into two files.
\begin{verbatim}
#ifndef EW_H
#define EW_H

#include <ginac/ginac.h>
#include <effective/effective.h>
#include <fstream.h>

const unsigned int SU2w = 2;
const unsigned int U1b = 1;

using namespace std;
using namespace GiNaC;
using namespace Effective;

const int famsize = 1;

class ElectroWeak : public Model {
public:
  ElectroWeak() : Model() {}
  virtual void createGaugeGroups();
  virtual void vreateGaugeFields();
  virtual void createMatterFields();
  virtual void addOtherTerms();
};

void ElectroWeak::createGaugeGroups() {
  addGaugeGroup(new U1Group("U1", "{g'}", this, U1b));
  addGaugeGroup(new SU2Group("SU2", "{g_W}", this, SU2w));
}

void ElectroWeak::createGaugeFields() {
  VectorSpin v;
  addField("B",new GaugeField("B","B",v,getGaugeGroup("U1"));
  addField("W",new GaugeField("W","W",v,getGaugeGroup("SU2"));
}

void ElectroWeak::createMatterFields() {
  numeric half(1,2);
  ScalarSpin s;
  FermionSpin f;
  GaugeGroup *u1 = getGaugeGroup("U1");
  GaugeGroup *su2 = getGaugeGroup("SU2");
  addField("l",  new MatterField("l", "\\ell", f, famsize,
                                 u1,-half,su2,1));
  addField("eR", new MatterField("eR","e_R",f,famsize,u1,-1));
  addField("H", new MatterField("H","H",s,1,u1,half,su2,1));

  // Now add Higgs Vev
  Parameter upsilon = addParameter("HiggsVev","upsilon",246.0,
                                   Parameter::vev);
  addVev("HiggsVev",getField("H"),lst(getIndex("H","SU2")==2),
         (upsilon+Model::star));
  addVevParameter(upsilon);
}

void ElectroWeak::addOtherTerms() {
  // Create mu^2 H H
  Parameter mu = addParameter("mu", "\\mu", 1.0);
  vector<idx*> indices = getField("H")->getIndices();
  ex H = getField("H")->expression();
  ex a = pow(mu,2)*H.conjugate()*H;
  add(Utils::sumIndices(a,indices).expand();
  
  // Create lambda * (HH)^2
  Parameter lambda = addParameter("lambda", "lambda", 1.0);
  ex la = Utils::sumIndices(H.conjugate()*H,indices).expand();
  add(-lambda*pow(la,2));

  // Now add lepton Yukawa
  ex Ye;
  if(famsize != 1) Ye = addFamilyMatrix("Ye","{Y^e}",famsize);
  else Ye = addParameter("Ye", "{Y^e}", 1.0);
  idx j = Utils::familyIndex(1,famsize);
  ex eR = getField("eR")->expession();
  ex l;
  if(famsize != 1) l = getField("l")->expression().conjugate()
            .subs(Utils::familyIndex(0,famsize)==j);
  else l = getField("l")->expression().conjugate();
  ex b = -Ye.subs(j==Utils::familyIndex(0,famsize))*l*H*eR;
  indices = getField("l")->getIndices();
  if(famsize != 1) indices.push_back(&j);
  ex res = Utils::sumIndices(b,indices).expand();
  add(2*Utils::real(res));
}
\end{verbatim}

\section{Summary of Classes and Routines}
This appendix provides a summary of some of the more important routines
discussed in this article.

\subsection{Classes and Routines Used for the Field Content}

In this section we present a summary of the classes and routines that are
needed to define the field content of a model.

Table \ref{tab:fcclass} gives a list of the classes encountered when defining
the field content. This is intended to give a brief synopsis of the relevant
classes one should use when defining ones field content.

\begin{table}[!htb]
  \begin{tabular}{ll}
    {\tmstrong{Class}} & {\tmstrong{Description}}\\ \hline
    \tmtexttt{GaugeGroup} & Abstract class defining the behaviour of a group\\
    \tmtexttt{U1Group} & The definition of the $U (1)$ group\\
    \tmtexttt{SU2Group} & The definition of the $S U (2)$ group\\
    \tmtexttt{SU3Group} & The definition of the $S U (3)$ group\\
    \tmtexttt{Spin} & Abstract class defining behaviour of spin objects\\
    \tmtexttt{ScalarSpin} & Definition of the scalar spin behaviour\\
    \tmtexttt{FermionSpin} & Definition of the fermion spin behaviour. This is
    specific\\
    & to the Weyl fermions. A new class is needed for Dirac\\
    & fermions\\
    \tmtexttt{VectorSpin} & Definition of the vector spin behaviour\\
    \tmtexttt{Field} & Abstract class of a set of fields with the same
    properties\\
    \tmtexttt{GaugeField} & Class specific to fields which mediate the
    interactions\\
    \tmtexttt{MatterField} & Class which describes the remaining fields\\
    \tmtexttt{Parameter} & Analytic object with a numeric value which can
    change
  \end{tabular}
  \caption{Table of the classes used to define the field
  content.\label{tab:fcclass}}
\end{table}

Table \ref{tab:fcrout} gives the list of routines which are used to define the
field content. The first column gives the routine; all routines in the table
are found in the \tmtexttt{Model} class. Information on the arguments to the
routines can be found in the text of the previous sections, or online at the
URL given in the first section.

\begin{table}[!htb]
  \begin{tabular}{ll}
    {\tmstrong{Routine}} & {\tmstrong{Description}}\\ \hline
    \tmtexttt{addGaugeGroup} & Adds a new group to the model\\
    \tmtexttt{addField} & Adds a field to the model\\
    \tmtexttt{addParameter} & Adds a new parameter to the model\\
    \tmtexttt{addVev} & Adds a vacuum expectation value to the model\\
    \tmtexttt{addVevParameter} & Tells model \tmtexttt{Parameter} is a VEV\\
    \tmtexttt{getGaugeGroup} & returns the pointer to the gauge group\\
    \tmtexttt{getField} & returns a pointer to a \tmtexttt{Field}\\
    \tmtexttt{getGaugeField} & returns a pointer to a \tmtexttt{GaugeField}\\
    \tmtexttt{getMatterField} & returns a pointer to a
    \tmtexttt{MatterField}\\
    \tmtexttt{getParam} & returns a \tmtexttt{Parameter} object\\
    \tmtexttt{getParameter} & returns a \tmtexttt{numeric} object
  \end{tabular}
  \caption{Routines used to define field content.\label{tab:fcrout}}
\end{table}

The \tmtexttt{Model} class is an abstract class. The user must define their
own concrete version of this class which contains their field content. Table
\ref{tab:fcmodel} describes the routines needed to define the field content of
of the concrete class.

\begin{table}[!htb]
  \begin{tabular}{ll}
    {\tmstrong{Routine}} & {\tmstrong{Description}}\\ \hline
    \tmtexttt{createGaugeGroups} & Routine where the \tmtexttt{GaugeGroup}
    objects are defined\\
    \tmtexttt{createGaugeFields} & Routine where the \tmtexttt{GaugeField}
    objects are defined\\
    \tmtexttt{createMatterFields} & Routine where the \tmtexttt{MatterField}
    objects are defined\\
    & and the VEV are given\\
    \tmtexttt{addOtherTerms} & Routine to define any other terms in the Lagrangian
  \end{tabular}
  \caption{Routines which are virtual in the \tmtexttt{Model} class and must
  be implemented by the user when defining their model.\label{tab:fcmodel}}
\end{table}

\subsection{Classes and Routines Used for the Interaction Terms}

Table \ref{tab:int} gives the list of routines in the \tmtexttt{Model} 
class that were used for creating the interaction terms.

\begin{table}[!htb]
  \begin{tabular}{lll}
    {\tmstrong{Class}} & {\tmstrong{Routine}} & {\tmstrong{Description}}\\ \hline
    \tmtexttt{Model} & \tmtexttt{addFamilyMatrix} & This creates a matrix with
    a unique \tmtexttt{Parameter}\\
    &  & for each individual element. One can retrieve\\
    &  & the whole matrix with \tmtexttt{getFamilyMatrix()}\\
    &  & or each element with \tmtexttt{getParameter()} or\\
    &  & \tmtexttt{getParam()}.\\
    \tmtexttt{Model} & \tmtexttt{getFamilyMatrix} & This returns the matrix
    (with unreferenced \\
    \tmtexttt{} &  & indices) given by the input string.\\
    \tmtexttt{Model} & \tmtexttt{add} & Add the expression to the
    Lagrangian.\\
    \tmtexttt{Field} & \tmtexttt{expression} & Returns the (complex)
    expression for a field.\\
    \tmtexttt{Field} & \tmtexttt{getIndices} & Returns all of the indices
    (except Lorentz)\\
    &  & for a \tmtexttt{Field}. There is an optional argument \\
    &  & which when \tmtexttt{false} removes the family index\\
    &  & from the list.\\
    \tmtexttt{Utils} & \tmtexttt{familyIndex} & Returns a symbolic index for
    the family space\\
    &  & from a predefined list of indices.\\
    \tmtexttt{Utils} & \tmtexttt{sumIndices} & Sums the expression over the
    indices given.
  \end{tabular}
  \caption{List of classes and routines used for the interaction terms.\label{tab:int}}
\end{table}

\subsection{Classes and Routines used for the Effective Potential}

Here we summarize the classes and routines which were needed
to deal with the effective potential. Table \ref{tab:numeric} gives the
descriptions of the two routines in the \tmtexttt{Numerics} class that were
encountered.

\begin{table}[!htb]
  \begin{tabular}{ll}
    {\tmstrong{Routine}} & {\tmstrong{Description}}\\ \hline
    \tmtexttt{extremizePotential} & This routine takes a list of
    \tmtexttt{Parameter}s and finds\\
    & the values which minimize the effective potential.\\
    & This routine also takes an approximation which to\\
    & use when evaluating the potential. This numerical\\
    & routine is based on the direction set (or Powell's) \\
    & method in multidimensions.\\
    \tmtexttt{solveZeroTadpoles} & This method is similar to the previous one
    except\\
    & it takes a list of \tmtexttt{Parameter}s and finds the values \\
    & which give tadpoles which are equal to 0. This\\
    & method can also take different approximations to \\
    & use to evaluate the tadpole diagrams. This method\\
    & uses the Newton-Raphson root finding method.
  \end{tabular}
  \caption{Routines in the \tmtexttt{Numerics} class used for the effective
  potential and tadpoles.\label{tab:numeric}}
\end{table}

Table \ref{tab:tadpot} shows the routines in the \tmtexttt{Model} class
which were used with the effective potential. We have also introduced the enumerated
type \tmtexttt{Approximation\-::\-Approximations} and two possible values it can take,
\tmtexttt{Approximation\-::TreeLevel} and \tmtexttt{OneLoop}.

\begin{table}[!htb]
  \begin{tabular}{ll}
    {\tmstrong{Routine}} & {\tmstrong{Description}}\\ \hline
    \tmtexttt{tadpole} & Returns the values of the tadpoles for the current
    values\\
    & of the parameters. This routine takes an argument which\\
    & specifies what approximation to use when evaluating the\\
    & tadpoles. If none is given the default result is the\\
    & tree-level tadpoles.\\
    \tmtexttt{potential} & Returns the effective potential for the current
    values of the\\
    & parameters. If the approximation given is tree-level (default)\\
    & then the result is returned analytically. If the approximation\\
    & is one-loop then it is returned numerically.
  \end{tabular}
  \caption{The two routines of the \tmtexttt{Model} class which were
    discussed in section~\ref{sec:effpot}.\label{tab:tadpot}}
\end{table}

\subsection{Classes and Routines for Masses and Mixing Angles}

Table \ref{tab:massma} shows the two constructors of the classes
\tmtexttt{Mass} and \tmtexttt{MixingAngle}. These classes are {\tmname{GiNaC}}
objects that can be used in expressions. The value of them is not evaluated
until a call to \tmtexttt{evalf()} is made. This means the user can create
expressions as functions of the masses and mixing angles of their model and
trust that the values will take the appropriate values each time
\tmtexttt{evalf()} is called. In order to make this process more efficient,
the masses aren't re-evaluated every time \tmtexttt{evalf()} is called. They
are re-evaluated if one of two conditions holds. The first is that the
approximation being used is different than the last one used. The second
condition is if a flag has been set to force re-evaluation. This is provided
by the function \tmtexttt{Model::resetMasses()}.

\begin{table}[!htb]
  \begin{tabular}{ll}
    {\tmstrong{Constructor}} & {\tmstrong{Description}}\\ \hline
    \tmtexttt{Mass(Model*,ex,MassCorrection)} & This constructor takes a
    \tmtexttt{Model} pointer\\
    & and a field and provides a {\tmname{GiNaC}}\\
    & object which evaluates the mass of the\\
    & field under the given approximation.\\
    \tmtexttt{MixingAngle(Model*,ex} & This constructor is similar to the
    \tmtexttt{Mass}\\
    \tmtexttt{ \ \ \ \ \ \ \ \ ex,MassCorrection)} & constructor except it
    takes two fields.\\
    & This provides the value for element \\
    & given by the two fields of the rotation\\
    & matrix used to diagonalize the mass\\
    & matrix.
  \end{tabular}
  \caption{Description of the \tmtexttt{Mass} and \tmtexttt{MixingAngle}
  constructors.\label{tab:massma}}
\end{table}

Table \ref{tab:mc} provides a summary of the four mass corrections provided
in \tmname{Effective}.

\begin{table}[!htb]
  \begin{tabular}{ll}
    {\tmstrong{Class}} & {\tmstrong{Description}}\\ \hline
    \tmtexttt{TreeLevel} & Simply diagonalizes the tree level mass matrix\\
    \tmtexttt{OneLoopMassApprox} & This uses the tree level diagonalized mass
    matrix\\
    & and computes corrections to the diagonal elements\\
    & only.\\
    \tmtexttt{OneLoopMass} & Solves, for $p^2$, the determinant equal to 0\\
    & in eqns.~(\ref{eqn:dets}) and~(\ref{eqn:detf}).\\
    \tmtexttt{OneLoop} & This uses the same approach as
    \tmtexttt{OneLoopMass}\\
    & and also computes the one-loop mixing angles.
  \end{tabular}
  \caption{The four default mass corrections provided in
  {\tmname{Effective}}.\label{tab:mc}}
\end{table}

\subsection{Classes and Routines for the RGEs}

Section~\ref{sec:rge} introduced a new class \tmtexttt{RGE}. This class was responsible
for properly treating the RGEs and the evolution of parameters between scales.
In this version of {\tmname{Effective}}, this class is simple and requires the
user to input the expressions for the $\beta$ functions. It is planned that in
future versions, an automatic one-loop calculation for all parameters will be
provided for at least one renormalization scheme.

Table \ref{tab:rge} shows a list of the routines of \tmtexttt{RGE} that have 
been described and used in the discussion of the RGEs.

\begin{table}[!htb]
  \begin{tabular}{ll}
    {\tmstrong{Routine}} & {\tmstrong{Description}}\\ \hline
    \tmtexttt{evolve} & This evolves the parameters from the current scale
    to\\
    & the scale provided.\\
    \tmtexttt{initialCondition} & This routine allows the user to define the
    default value\\
    & of a parameter at a particular scale. None of the\\
    & initial conditions need be defined at the same scale.\\
    \tmtexttt{applyInitial} & This routine takes a scale, $\mu$, and a
    $\delta$ value as \\
    & arguments. It then finds all parameters with initial \\
    & conditions defined at the $\mu \pm \delta$ and applies them.\\
    \tmtexttt{print} & This prints the analytic form of the $\beta$ functions
    to the\\
    & stream provided.
  \end{tabular}
  \caption{The list of routines of the class \tmtexttt{RGE}.\label{tab:rge}}
\end{table}

\section{Effective Potential Review}

We will begin by deriving the effective potential. We will then show how the
one-loop effective potential can be derived in a model-independent way. This
means that the one-loop correction only depends on the masses of the model,
not explicitly on the couplings. This review is derived from~\cite{Quiros}. The
interested reader can find a more thorough discussion there.

We begin by considering a theory of one scalar field $\phi$ with a Lagrangian
density $\mathcal{L}\{\phi (x)\}$. The action is given by
\begin{eqnarray}
  S [\phi] & = & \int d^4 x\mathcal{L}\{\phi\}. 
\end{eqnarray}
The vacuum-to-vacuum expectation value $\left\langle 0_{\tmop{out}}
|0_{\tmop{in}} \right\rangle_j$ is given by
\begin{eqnarray}
  Z [j] & = & \left\langle 0_{\tmop{out}} |0_{\tmop{in}} \right\rangle_j
  \equiv \int \mathcal{D} \phi \exp \{i (S [\phi] + \phi j)\}, 
  \label{eqn:v2v}
\end{eqnarray}
where
\begin{eqnarray}
  \phi j & = & \int d^4 x \phi (x) j (x) . 
\end{eqnarray}
We can define the connected generating functional, $W [j]$, in terms of the
vacuum-to-vacuum expectation value
\begin{eqnarray}
  Z [j] & \equiv \exp \{i W [j]\}. &  \label{eqn:gf}
\end{eqnarray}
We now define the effective action, $\Gamma [\phi]$, for $S [\phi]$ such that
its classical field equation is the solution to the Schwinger-Dyson equation
for $S [\phi]$. That is, we require $\Gamma' [\phi] = j$. Solving this equation 
gives
\begin{eqnarray}
  \Gamma [ \bar{\phi}] & = & W [j] - \int d^4 x \frac{\delta W [j]}{\delta j
  (x)} j (x), 
\end{eqnarray}
where
\begin{eqnarray}
  \bar{\phi} (x) & = & \frac{\delta W [j]}{\delta j (x)} . 
\end{eqnarray}
$\bar{\phi} (x)$ is then a weighted average of the fluctuations of the field
$\phi$. In a translationally invariant theory, this is a constant.
{\tmname{Effective}} is designed to only deal with translationally invariant
theories, therefore, $\overline{\phi_{}} (x)$ must be a constant, $\phi_c$,
which is the VEV of the field. The effective potential can then be defined as
\begin{eqnarray}
  \Gamma [\phi_c] & = & - \int d^4 x V_{\tmop{eff}} (\phi_c) . 
\end{eqnarray}
We now write this as an expansion around $\phi_c = 0$
\begin{eqnarray}
  V_{\tmop{eff}} (\phi_c) & = & - \sum_{n = 0}^{\infty} \frac{1}{n!} \phi_c^n
  \Gamma^{(n)} (p_i = 0),  \label{eqn:effp}
\end{eqnarray}
where the $\Gamma^{(n)}$ are the one-particle irreducible (1PI) Green
functions. If we now minimize the potential over the constant field $\phi_c$
we find the vacuum state of the theory.

At tree level, the effective potential, eqn. (\ref{eqn:effp}), is identical to
the classical effective potential. This can be stated simply as
\begin{eqnarray}
  V_{\tmop{tree}} & = & -\mathcal{L}(\phi_i \rightarrow \phi_{i, c}) . 
\end{eqnarray}
We now discuss the one-loop correction to this potential for a model with one
self-interacting scalar field. The results will generalize to all fields. This
simple model is given by
\begin{eqnarray}
  \mathcal{L} & = & \frac{1}{2} \partial^{\mu} \phi \partial_{\mu} \phi -
  \frac{1}{2} m^2 \phi^2 - \frac{\lambda}{4!} \phi^4 . 
\end{eqnarray}
As just shown, the one-loop correction to the tree-level effective potential
is given by the sum of all 1PI diagrams with a single loop and zero external
momenta. The $n$th diagram has $n$ propagators, $n$ vertices and $2 n$
external legs. The propagators contribute a factor of $i^n (p^2 - m^2 - i
\epsilon)^{- n}$. Each pair of external lines contributes a factor $\phi^{2
n}_c$ and each vertex a factor $- i \lambda / 2$. Including a global symmetry
factor we have
\begin{eqnarray}
  V_1 (\phi_c) & = & i \sum_{n = 1}^{\infty} \int \frac{d^4 p}{(2 \pi)^4}
  \frac{1}{2 n} \left[ \frac{\lambda \phi_c / 2}{p^2 - m^2 - i \epsilon}
  \right]^n \nonumber\\
  & = & - \frac{i}{2} \int  \frac{d^4 p}{(2 \pi)^4} \log \left[ 1 -
  \frac{\lambda \phi^2_c / 2}{p^2 - m^2 - i \epsilon} \right] . 
\end{eqnarray}
After a Wick rotation in the $\overline{\tmop{DR}}$ scheme~\cite{DRscheme} this is
\begin{eqnarray}
  V_1 (\phi_c) & = & \frac{1}{64 \pi^2} m^4 (\phi_c) \left( \ln \frac{m^2
  (\phi_c)}{\mu^2} - \frac{3}{2} \right), 
\end{eqnarray}
where
\begin{eqnarray}
  m^2 (\phi_c) & = & \frac{d^2 V_0 (\phi_c)}{d \phi_c^2}, 
\end{eqnarray}
is the tree-level mass and $\mu$ is the renormalization scale.

\section{Passarino Veltman Functions}

In this appendix we provide the definitions of the Passarino-Veltman
functions~\cite{Passarino:1979jh} that are provided in {\tmname{Effective}}. These can be
accessed by calling, for example, \tmtexttt{Passarino\_Veltman::A0()}. The
derivatives can be found by calling the {\tmname{GiNaC}}
\tmtexttt{diff()} routine.
\begin{eqnarray*}
  A_0 (m^2) & = & \frac{16 \pi^2}{\mu^{d - 4}}  \int  \frac{i d^d q}{(2
  \pi)^d} \frac{1}{q^2 - m^2 + i \epsilon},\\
  B_0 (p^2, m_1^2, m_2^2) & = & \frac{16 \pi^2}{\mu^{d - 4}}  \int  \frac{i
  d^d q}{(2 \pi)^d} \frac{1}{\left( q^2 - m_1^2 + i \epsilon \right) \left( (q
  + p)^2 - m_2^2 + i \epsilon \right)},\\
  p_{\mu} B_1 (p^2, m_1^2, m_2^2) & = & \frac{16 \pi^2}{\mu^{d - 4}}  \int 
  \frac{i d^d q}{(2 \pi)^d} \frac{q_{\mu}}{\left( q^2 - m_1^2 + i \epsilon
  \right) \left( (q + p)^2 - m_2^2 + i \epsilon \right)},\\
  g_{\mu \nu} B_{00}(p^2, m_1^2, m_2^2) &+& 
  p_{\mu} p_{\nu} B_{11} (p^2, m_1^2, m_2^2)\\
  & = & \frac{16 \pi^2}{\mu^{d - 4}}  \int  \frac{i d^d q}{(2 \pi)^d} \frac{q_{\mu}
  q_{\nu}}{\left( q^2 - m_1^2 + i \epsilon \right) \left( (q + p)^2 - m_2^2 +
  i \epsilon \right)} .
\end{eqnarray*}
The first two, $A_0$ and $B_0,$ can be expressed to $\mathcal{O}(\varepsilon)$
for $d = 4 - 2 \varepsilon$ as
\begin{eqnarray*}
  A_0 (m^2 ; \mu^2) & = & m^2 \left( \frac{1}{\bar{\varepsilon}} - 1 + \ln
  \frac{m^2}{\mu^2} \right) +\mathcal{O}(\varepsilon),\\
  B_0 (p^2, m_1^2, m_2^2 ; \mu^2) & = & \frac{1}{\bar{\varepsilon}} - \ln
  \frac{p^2}{\mu^2} - f_B (x_+) - f_B (x_-) +\mathcal{O}(\varepsilon) .
\end{eqnarray*}
In {\tmname{Effective}} we subtract only the terms proportional to
\begin{eqnarray*}
  \frac{1}{\bar{\varepsilon}} & = & \frac{1}{\varepsilon} - \gamma_E + \ln 4
  \pi .
\end{eqnarray*}
We also have
\begin{eqnarray*}
  x_{\pm} = \frac{s \pm \sqrt{s^2 - 4 p^2 (m_1^2 + i \epsilon}}{2 p^2} & , &
  f_B (x) = \ln (1 - x) - x \ln \left( 1 - \frac{1}{x} \right) - 1,
\end{eqnarray*}
and $s = p^2 - m_2^2 + m_1^2$.

\section{Mass Corrections}
In this section we provide the formula for all of the model independent
one-loop mass corrections.

\subsection{Scalar-Scalar}
\label{sec:ss}

We now define the contributions to the two-point Green functions for
scalar fields. There are 6 classes of diagrams which can contribute to the
full two-point Green function for scalar fields: two four-point
diagrams and four three-point diagrams.

The four-point diagrams have an internal field as a vector or a scalar. We
designate these two contributions as $\Pi^{S 4}$ and $\Pi^{V 4}$. It is
assumed that the scalar four point coupling is a simple scalar quantity, $C_{S
4}$, where the vector four-point coupling is a tensor of the form $C_{V 4}
g^{\mu \nu}$. The contributions are then
\begin{eqnarray}
  \Pi^{S 4} (p^2, m^2 ; \mu^2) & = & \frac{C_{S 4}}{16 \pi^2} A_0 (m^2 ;
  \mu^2), \\
  \Pi^{V 4} (p^2, m^2 ; \mu^2) & = & - \frac{d C_{V 4}}{16 \pi^2} A_0 (m^2 ;
  \mu^2) . 
\end{eqnarray}
The spin of the internal fields for the three-point diagrams can be pairs of
all three types of spins, as well as a vector-scalar pair. We designate these
contributions as $\Pi^{S 3}, \Pi^{F 3}, \Pi^{V 3}$ and $\Pi^{\tmop{VS}}$. We
assume that the scalar and fermion three-point couplings are simple scalar
quantities, $C_{S 3}$ and $C_{F 3}$ respectively. This leads to the
contributions
\begin{eqnarray}
  \Pi^{S 3} (p^2, m_1^2, m_2^2 ; \mu^2) & = & \frac{C_{S 3}^{(1)} C_{S
  3}^{(2)} }{16 \pi^2} B_0 (p^2, m_1^2, m_2^2 ; \mu^2), \\
  \Pi^{F 3} (p^2, m_1^2, m_2^2 ; \mu^2) & = & \frac{d C_{F 3}^{(1)} C_{F
  3}^{(2)}}{16 \pi^2} \left[ p^2 (B_1 (p^2, m_1^2, m_2^2 ; \mu^2) \right.
  \nonumber\\
  &  & + B_{11} (p^2, m_1^2, m_2^2 ; \mu^2) + d B_{00} (p^2, m_1^2, m_2^2 ;
  \mu^2) \nonumber\\
  &  & \left. + m_1 m_2 B_0 (p^2, m_1^2, m_2^2 ; \mu^2) \right], 
\end{eqnarray}
where the superscripts differentiate between the two vertices in the
three-point diagram.

The vector three-point coupling is assumed to have the Lorentz structure
$C_{V 3} g^{\mu \nu}$. Again, using the Feynman gauge, this gives
\begin{eqnarray}
  \Pi^{V 3} (p^2, m_1^2, m_2^2 ; \mu^2) & = & \frac{d C_{V 3}^{(1)} C_{V
  3}^{(2)}}{16 \pi^2} B_0 (p^2, m_1^2, m_2^2 ; \mu^2) . 
\end{eqnarray}
Lastly we turn to the scalar-vector three point coupling. It is assumed that
these couplings are derived from terms in the Lagrangian $C_{\tmop{VS}}
\varphi_j \partial_{\mu} \varphi_i A_{\nu} g^{\mu \nu}$. This means we must
consider the derivative to lie on both the external and internal fields. Since
we have two couplings of this type we find three contributions: the derivative
is on both of the external fields, the derivative is on both the internal
fields, and the derivatives lie one on the external and one on the internal.
These are denoted by $\Pi^{\tmop{VS}}_E, \Pi^{\tmop{VS}}_I$ and
$\Pi^{\tmop{VS}}_{I E}$, respectively, and are given by
\begin{eqnarray}
  \Pi^{V S}_E (p^2, m_V^2, m_S^2 ; \mu^2) & = & \frac{C_{\tmop{VS}}^{(1)}
  C_{\tmop{VS}}^{(2)}}{16 \pi^2} p^2 B_0 (p^2, m_V^2, m_S^2 ; \mu^2) 
  \label{eqn:pivse}\\
  \Pi^{V S}_I (p^2, m_V^2, m_S^2 ; \mu^2) & = & \frac{C_{\tmop{VS}}^{(1)}
  C_{\tmop{VS}}^{(2)}}{16 \pi^2} \left[ p^2 B_{11} (p^2, m_V^2, m_S^2 ;
  \nonesep \mu^2) \right. \nonumber\\
  &  & \left. + d B_{00} (p^2, m_1^2, m_2^2 ; \mu^2) \right] \\
  \Pi^{V S}_{I E} (p^2, m_V^2, m_S^2 ; \mu^2) & = & \frac{C_{\tmop{VS}}^{(1)}
  C_{\tmop{VS}}^{(2)}}{16 \pi^2} p^2 B_1 (p^2, m_V^2, m_S^2 ; \mu^2) . 
  \label{eqn:pivsie}
\end{eqnarray}
To get the total contribution for the vector-scalar three point coupling we
must sum these terms, i.e. $\Pi^{V S} = \Pi^{V S}_I + \Pi^{V S}_E + 2 \Pi^{V
S}_{I E}$, where the factor of 2 is from the two combinations of placing the
derivative on the internal and external fields. It is also worth noting that
the couplings between the three contributions in eqn.
(\ref{eqn:pivse}-\ref{eqn:pivsie}) do not have to be equal.

In order to get the complete scalar-scalar two-point Green function we must
sum over all possible internal fields for each contribution. This is
\begin{eqnarray}
  \Pi_{k l} (p^2) & = & \sum_{i \in \tmop{vector}} \left[ \Pi^{V 4}_i +
  \sum_{j \in \tmop{scalar}} \Pi_{i j}^{V S} + \sum_{j \in \tmop{vector}}
  \Pi_{i j}^{V 3} \right] + \sum_{i, j \in \tmop{fermion}} \Pi^{F 3}_{i j}
  \nonumber\\
  &  & + \sum_{i \in \tmop{scalar}} \left[ \Pi^{S 4}_i + \sum_{j \in
  \tmop{scalar}} \Pi_{i j}^{S 3} \right], 
\end{eqnarray}
where the indices $k$ and $l$ represent the external scalar fields and indices
$i$ and $j$ represent the choice of internal loop fields. Of course, in this
sum many choices of $i$ and $j$ do not contribute as these fields don't couple
to $k$ and $l$.

\subsection{Fermion-Fermion}
\label{sec:ff}

The fermion-fermion self energy consists only of three point coupling diagrams
with a a fermion-vector or a fermion-scalar internal loop pair. The
scalar-fermion vertex is assumed to be a simple scalar coupling, $C_S$, while
the vector-fermion coupling is assumed to be of the form $C_V \gamma^{\mu}$.
The contribution from these diagrams \ are
\begin{eqnarray}
  \Sigma^S ( \not{p}, m_f^2, m_S^2 ; \mu^2) & = & \frac{C_S^{(1)}
  C_S^{(2)}}{16 \pi^2} \left\{ \not{p} B_1 (p^2, m_f^2, m_S^2 ; \mu^2) \right. \nonumber\\
  && \left. + m_f  B_0 (p^2, m_f^2, m_S^2, \mu^2) \right\} \\
  \Sigma^V ( \not{p}, m_f^2, m_V^2 ; \mu^2) & = & \frac{C_V^{(1)}
  C_V^{(2)}}{16 \pi^2} \left\{ (d - 2) \not{p} B_1 (p^2, m_f^2, m_V^2 ; \mu^2)
  \right.\nonumber \\
  && \left. - d \, m_f B_0 (p^2, m_f^2, m_V^2 ; \mu^2) \right\} . 
\end{eqnarray}
Again to construct the full two-point Green function one must sum over all
possible choices of internal fields. The two-point Green function for fields
$k$ and $l$ is
\begin{eqnarray}
  \Sigma_{k l} ( \not{p}) & = & \sum_{i \in \tmop{fermion}} \left[ \sum_{j \in
  \tmop{scalar}} \Sigma^S_{i j} + \sum_{j \in \tmop{vector}} \Sigma^V_{i j}
  \right] . 
\end{eqnarray}

\subsection{Vector-Vector}
\label{sec:vv}

The vector-vector two-point Green function is composed of seven classes of
diagrams, two four-point coupling diagrams and five three-point coupling
diagrams.

The four-point couplings diagrams contain either an internal scalar or vector
field. The contributions from these diagrams are denote by $\Pi_{\mu \nu}^{S
4}$ and $\Pi_{\mu \nu}^{V 4}$, respectively. The scalar four-point coupling
between scalar $\varphi$ and vector fields $A^{\mu}_a A^{\nu}_b$ is assumed to
have the form $C_{S 4} g^{\mu \nu}$, where $C_{S 4}$ implicitly contains the
$a$ and $b$ indices. The contribution is
\begin{eqnarray}
  \Pi_{\mu \nu}^{S 4} (p^2, m^2 ; \mu^2) & = & \frac{C_{S 4}}{16 \pi^2} g_{\mu
  \nu} A_0 (m^2 ; \mu^2) . 
\end{eqnarray}
The vector four-point coupling between vectors $A^{\mu}_a, A^{\nu}_b,
A^{\lambda}_c$ and $A^{\sigma}_d$ is given by $C_{V 4}^{(1)} g^{\mu \nu}
g^{\lambda \sigma} + C_{V 4}^{(2)} g^{\mu \lambda} g^{\nu \sigma} + C_{V
4}^{(3)} g^{\mu \sigma} g^{\nu \lambda}$ with the group indices $a, b, c$ and
$d$ contained in the coefficients $C_{V 4}^{(1, 2, 3)}$. The contribution from
the diagram is
\begin{eqnarray}
  \Pi_{\mu \nu}^{V 4} (p^2, m^2 ; \mu^2) & = & - \frac{d C_{V 4}^{(1)} + C_{V
  4}^{(2)} + C_{V 4}^{(3)}}{16 \pi^2} g_{\mu \nu} A_0 (m^2 ; \mu^2) . 
\end{eqnarray}
This result is found for $d$ dimensions and in the Feynman gauge.

We now turn to the contributions from the three-point coupling diagrams. We
begin with the diagram which contains a pair of scalars for the internal
fields. The couplings for this diagram are assumed to come from a term $C_{S
3} \varphi_j \partial_{\mu} \varphi_i A^{\mu}$. As was seen for the
scalar-scalar two-point Green function we must consider the placement of the
derivative for both of the vertices. This means we have four possible
contributions which we denote by $\Pi_{\mu \nu}^{S 3 (i, j)}$, for $i,j = 1,2$. We have the
first contribution when the derivative is associated with the first loop field
(the one associated with $m_1$) in both vertices. This is given by
\begin{eqnarray}
  \Pi^{S 3 (1, 1)}_{\mu \nu} (p^2, m_1^2, m_2^2 ; \mu^2) & = & - \frac{C_{S
  3}^{(1, 1)} C_{S 3}^{(2, 1)}}{16 \pi^2}  \left[ g_{\mu \nu} B_{00} (p^2,
  m_1^2, m_2^2 ; \mu^2) \right. \nonumber\\
  &  & + \left. p_{\mu} p_{\nu} B_{11} (p^2, m_1^2, m_2^2 ; \mu^2) \right] 
\end{eqnarray}
where $C_{S 3}^{(i, j)}$ represent the coefficient of the term for the $i$th
vertex where the derivative is placed on the $j$th field in the loop.
Similarly we find for the remaining contributions
\begin{eqnarray}
  \Pi_{\mu \nu}^{S 3 (2, 2)} (p^2, m_1^2, m_2^2 ; \mu^2) & = & - \frac{C_{S
  3}^{(1, 2)} C_{S 3}^{(2, 2)}}{16 \pi^2} \left( g_{\mu \nu} B_{00} (p^2,
  m_1^2, m_2^2 ; \mu^2) \right. \nonumber\\
  &  & \left. + p_{\mu} p_{\nu}  B_{11} (p^2, m_2^2, m_1^2 ; \mu^2) \right), \\
  \Pi_{\mu \nu}^{S 3 (1, 2)} (p^2, m_1^2, m_2^2 ; \mu^2) & = & \frac{C_{S
  3}^{(1, 1)} C_{S 3}^{(2, 2)}}{16 \pi^2} \left( g_{\mu \nu} B_{00} (p^2,
  m_1^2, m_2^2 ; \mu^2) \right. \nonumber\\
  &  & \left. + p_{\mu} p_{\nu}  \left[ \left( B_1 + B_{11} \right) (p^2,
  m_1^2, m_2^2 ; \mu^2) \right] \right), \\
  \Pi_{\mu \nu}^{S 3 (2, 1)} (p^2, m_1^2, m_2^2 ; \mu^2) & = & \frac{C_{S
  3}^{(1, 2)} C_{S 3}^{(2, 1)}}{16 \pi^2} \left( g_{\mu \nu} B_{00} (p^2,
  m_1^2, m_2^2 ; \mu^2) \right. \nonumber\\
  &  & \left. + p_{\mu} p_{\nu}  \left[ \left( B_1 + B_{11} \right) (p^2,
  m_1^2, m_2^2 ; \mu^2) \right] \right) . 
\end{eqnarray}
In {\tmname{Effective}} it also assumed that there is a symmetry between the
terms of $\varphi_j \partial_{\mu} \varphi_i A^{\mu}$ and $\varphi_i
\partial_{\mu} \varphi_j A^{\mu}$ such that $C_{S 3}^{(k,1)} = C_{S 3}^{(k,2)}$
for $k = 1, 2$. This means all four terms have the same coefficient and
can be added. Doing so one find the result
\begin{eqnarray}
  \Pi^{S 3}_{\mu \nu} (p^2, m_1^2, m_2^2 ; \mu^2) & = & - \frac{C_{S 3}^{(1)}
  C_{S 3}^{(2)}}{16 \pi^2} p_{\mu} p_{\nu} B_0 (p^2, m_1^2, m_2^2 ; \mu^2) . 
\end{eqnarray}
We now turn to the fermion three-point coupling and the vector-scalar three
point coupling. In the fermion case the coupling is assumed to take the form
$C_{F 3} g^{\mu \nu} \gamma_{\mu} \psi_i \psi_j A_{\nu} $ where $C_{F 3}$ is a
scalar coefficient. The vector-scalar coupling is assumed to be $C_{V S}
g^{\mu \nu} \varphi_i A_{\mu} A_{\nu}$, with $C_{V S}$ a scalar. The
contributions from these diagrams is then
\begin{eqnarray}
  \Pi_{\mu \nu}^{F 3} (p^2, m_1^2, m_2^2 ; \mu^2) & = & \frac{d C_{F 3}^{(1)}
  C_{F 3}^{(2)}}{16 \pi^2} \left[ 2 p_{\mu} p_{\nu} \left( B_{11} - B_1
  \right) - g_{\mu \nu} \left( m_1 m_2 B_0 \right. \right. \nonumber\\
  &  & \left. + \left. (d - 2) B_{00} - p^2 (B_{11} - B_1) \right) \right],
  \\
  \Pi_{\mu \nu}^{V S} (p^2, m_1^2, m_2^2 ; \mu^2) & = & -_{} \frac{C_{V
  S}^{(1)} C_{V S}^{(2)}}{16 \pi^2} g_{\mu \nu} B_0 (p^2, m_1^2, m_2^2 ;
  \mu^2), 
\end{eqnarray}
where the arguments to the functions in the fermion contribution have been
suppressed for simplicity.

We then have the vector-vector three-point coupling. It is
assumed that these are derived from terms like $\partial_{\mu} A_{\nu}
A_{\lambda} A_{\sigma} (C_1 g^{\mu \lambda} g^{\nu \sigma} + C_2 g^{\mu
\sigma} g^{\nu \lambda})$. When the derivative is permutated to lie on each
possible vector field, this gives rise to 6 scalar coefficients per vertex and
9 diagrams. Symmetries reduce this to only 5 independent contributions. Due to
the large number of contributions, these will not be listed here, but the sum
will be denoted by $\Pi_{\mu \nu}^{V 3}$.

Lastly, because the previous results were computed in the Feynman gauge, we
must also include the three-point diagram where ghost fields are the intermediate
fields. Contributions of this sort give a contribution
\begin{eqnarray}
\Pi_{\mu \nu}^G (p^2,m_{G1}^2, m_{G2}^2 ; \mu^2) &=& \frac{C^{(1)} C^{(2)}}{16 \pi^2} 
\left( g_{\mu \nu} B_{00} + p_\mu p\nu \left[ B_1 - B_{11} \right] \right),
\end{eqnarray}
where again, we have suppressed the arguments of the functions on the right-hand
side of the equation.

As only the transverse components of the vector field propagate, only the
transverse component of the Green function needs to be used in
renormalization. This means that only the coefficient of the $g_{\mu \nu}$
terms are used in the one-loop mass corrections. We denote this component for
the Green function between vector fields $k$ and $l$ as $\Pi^T_{k l}$. Thus
\begin{eqnarray}
  \Pi^T_{k l} (p^2) & = & \sum_{i \in \tmop{scalar}} \left( \Pi^{T, S 4}_i +
  \sum_{j \in \tmop{scalar}} \Pi^{T, S 3}_{i j} + \sum_{j \in \tmop{vector}}
  \Pi^{T, V S}_{i j} \right) \nonumber\\
  &  & + \sum_{i, j \in \tmop{fermion}} \Pi^{T, F 3}_{i j} + \sum_{i \in
  \tmop{vector}} \left( \Pi^{T, V 4}_i + \sum_{j \in \tmop{vector}} \Pi^{T, V
  3}_{i j} \right) \nonumber \\
  & & + \sum_{i,j \in \tmop{ghost}} \Pi^{T,G}_{ij}, 
\end{eqnarray}
where again the indices $k$ and $l$ are implicit in the couplings of the terms
on the right-hand side. In \tmname{Effective} the ghost fields have the mass
of the boson field with the same group charge. The ghost-ghost-vector three 
point coupling is assumed to be of the form $g f^{abc}$. 

\bibliography{effective}
\bibliographystyle{JHEP-2}
\end{document}